\documentclass[12pt]{article}
\usepackage{amsmath,amsfonts}
\usepackage{amssymb,latexsym}
\usepackage{epsfig}
\usepackage{graphicx}
\usepackage{subfigure}
\usepackage{cite}

\newtheorem{theorem}{Theorem}

\newtheorem{defi}{Definition}[section]

\def\R{\mathbb{R}}
\def\N{\mathbb{N}}
\def\C{\mathbb{C}}
\def\Z{\mathbb{Z}}

\def\H{\mathcal{H}}

\newcommand{\G}{\Gamma}

\newcommand{\p}{\mathbf{p}}
\newcommand{\q}{\mathbf{q}}

\def\sd{{}^{[s]}D_q}

\def\qn{{}^{[s]}[n]_q}
\def\qnp{{}^{[s]}[n+1]_q}
\def\qnm{{}^{[s]}[n-1]_q}

\def\sm{{}^{[s]}d_q}

\def\mfe{\mathfrak{e}}
\def\mfE{\mathfrak{E}}

\def\lg{\langle }
\def\rg{\rangle }

\def\deq{:=}

\begin{document}


\thispagestyle{empty}
\hfill \today

\vspace{2.5cm}

\begin{center}
\bf{\LARGE Pisot $q$-Coherent states quantization \\[0.4cm]
of the harmonic oscillator}
\end{center}

\vskip1.25cm

\begin{center}
J.P. Gazeau$^1$ and M.A. del Olmo$^2$
\end{center}

\begin{center}
$^1${\sl Laboratoire APC, Univ Paris Diderot, Sorbonne Paris Cit\'e,\\
 75205 Paris-Fr}\\
\medskip

$^2${\sl Departamento de F\'{\i}sica Te\'orica, Universidad de
Valladolid, \\
E-47005, Valladolid, Spain.}\\
\medskip

{e-mail: gazeau@apc.univ-paris7.fr , olmo@fta.uva.es}

\end{center}
\vskip1.25cm

\begin{abstract}
We revisit the quantized version of the harmonic oscillator obtained through a $q$-dependent family of coherent states. For  each $q$, $0< q < 1$,  these normalized states form an overcomplete set that  resolves the unity with respect to an explicit  measure. We restrict our study to the case in which $q^{-1}$ is a quadratic unit Pisot number: the  $q$-deformed integers form Fibonacci-like  sequences of integers.   We then examine the main characteristics of the corresponding quantum oscillator: localization in the configuration and in the phase spaces,  angle operator, probability distributions and related statistical features,  time evolution  and  semi-classical phase space trajectories.  

\bigskip


\smallskip


\end{abstract}

\section{Introduction}
\label{intro}
In this work we revisit non-linear coherent states obtained from the standard ones through a certain type of $q$-- or $q\,p$--deformations of the integers. Similar deformations have already been considered during the two last decades, since the pioneering work by Arik \emph{et al} \cite{arik92} (see for instance \cite{quesne02,curado1,gavrilik1,gavrilik2} and references therein for a list of more recent related works, mostly devoted to algebraic aspects of such deformations).  Our aim is to use  these deformed coherent states  for quantizing  \`a la Berezin-Klauder-Toeplitz elementary classical observables like position, momentum, angle, quadratic Hamiltonian etc,  and to examine some interesting features of   their quantum counterparts.   In particular, we insist on the condition that our  (symmetric) $q$-deformations of integers be still integers, more precisely yield sequences of integers that generalize the famous Fibonacci sequence. 

Section~\ref{csquant} is a review of what we understand by quantization of a centered disk of finite radius  or of the complex plane $\C$, both being viewed as possible phase spaces for one degree of freedom mechanical systems.    
We then study  in  Section~\ref{Integersasdeformationsofintegers} sequences of integer numbers that appear as $q\,p$-deformations of integer numbers.  We present two types of such sequences that we call fermionic (or antisymmetric) and bosonic (or symmetric) sequences. In both cases $p$ and $q$ are  quadratic Pisot (more fairly Pisot-Vijayaraghavan) numbers.  Among these  appears the  well-known Fibonacci   sequence. 
We specially concentrate our study upon sequences of integers $({}^s[n]_q)_{n\in \N}$ encountered in symmetric $q$-deformations of integers. 
Section~\ref{momentumproblem} is devoted to the resolution of the moment problem \eqref{genmom}, i.e, to finding a positive  measure $w_q(t)\,dt$ such that ${}^s[n]_q != \int_0^{\infty} t^n\, w_q(t) \, dt $. 
In Section~\ref{CSqsymmetricdeformations} we present some computations and graphics for different $q$-sequences showing the main physical properties of these $q$-CS and the exceptional Pisot cases leading to nice periodic phase trajectories for lower symbols of time-evolving  position and momentum operators issued from CS quantization. 
In the conclusion we give some  remarks and put our work into the perspective of the general coherent state quantization framework.  
Finally four appendices are devoted to an introduction to the standard $q$-calculus, the symmetric $q$-calculus, the equations determining the quadratic Pisot numbers and powers and recurrence for general quadratic Pisot numbers.

\section{A review of non-linear CS quantization issued from sequences of numbers}
\label{csquant}
 Let us suppose that we observe through some experimental device an infinite, strictly increasing sequence of
nonnegative real numbers
$\left( x_n\right)_{n\in \N}$,  such that $x_0 = 0$, for instance the  quantum energy spectrum of a given system, but it could be some other kind of observed data. 
To this sequence of numbers there corresponds the sequence of ``factorials''
$x_n! := x_1\times x_2\times \dotso \times x_n$ with  $x_0! := 1$.
An  associated exponential can be defined by
 \begin{equation}
\label{seqexp}
\mathcal{N}(t): = \sum_{n = 0}^{+\infty} \frac{t^n}{x_n!} \, .
\end{equation}
We assume that it has a nonzero  convergence radius $R = \lim_{n \to \infty} 
x_{n+1}$.
We now suppose that the (Stieltjes) moment problem has a (possibly non unique)
solution for the sequence of factorials $\left(x_n!\right)_{n\in \N}$, i.e.
there exists a probability distribution $t \mapsto w(t)$  on $[0, R)$
such that
\begin{equation}
\label{genmom}
x_n! = \int_{0}^{R} t^n \, w(t)\, dt\, ,
\end{equation}
and we extend (formally)  the definition of the function $n\mapsto x_n!$ to real or complex numbers when it is needed (and possible!).
Let $\H$ be a separable (complex) Hilbert space with scalar
product $\lg\cdot |\cdot\rg$  and orthonormal basis
$|e_0\rg,|e_1\rg,\dotsc, | e_n\rg, \dotsc$, i.e.,
\begin{equation}
\label{scprod}
\lg e_n|e_{n'}\rg = \delta_{nn'}\,, \qquad 
\sum_{n \geq 0} |e_n\rg\lg e_n| = I_d .
\end{equation}
To each complex number $z$ in  the open disk 
$\mathcal{D}_{\sqrt{R}} = \{z \in \C\, , \, \vert z \vert <  \sqrt{R} \}$ we associate a  vector in $\H$ defined as:
\begin{equation}
\label{xncs}
|v_z\rg = \sum_{n=0}^{\infty} \frac{1}{\sqrt{\mathcal{N}(\vert z \vert^2)}}\frac{z^n}{\sqrt{x_n!}}\,  |e_n\rg\, .
\end{equation}
These vectors  enjoy the following  properties similar to those obeyed by standard  coherent states (CS) \cite{klauskag,gazeau09}:
\begin{itemize}
  \item[(i)] $\Vert v_z\Vert = 1$ (\textit{normalization}).
  \item[(ii)] The map $\mathcal{D}_{\sqrt{R}} \ni z \mapsto |v_z\rg $ is weakly
  continuous (\textit{continuity}).
  \item[(iii)] The map $\N \in n \mapsto \vert \lg e_n |v_z\rg \vert^2 =
  \dfrac{\vert z \vert^{2n}}{\mathcal{N}(\vert z \vert^2)\;x_n!}$ is a
  Poisson-like distribution in $\vert z\vert^2$,  with average number of events equal
 to $\vert z \vert^2$
  (\textit{discrete probabilistic content}).
\item[(iv)] The map $\mathcal{D}_{\sqrt{R}}  \ni z \mapsto \vert \lg e_n |v_z\rg
\vert^2 = \dfrac{\vert z \vert^{2n}}{\mathcal{N}(\vert z \vert^2)\;
x_n!}$ 
{is} a (gamma-like) probability distribution (with
respect to the square of the radial variable) with $x_{n+1}$ as a
shape parameter 
{and} with respect to the following measure on the open disk 
$\mathcal{D}_{\sqrt{R}}$ (\textit{continuous
probabilistic content} \cite{alhelgaz})
\begin{equation}
\label{modmeas}
\nu(d^2z)
:= \mathcal{N}(\vert z \vert^2)\, w(\vert z\vert^2) \frac{d^2 z}{ \pi}\, ,
\end{equation}
where $d^2 z=d\,{\rm Re} z\;d\,{\rm Im} z$.

\item[(v)] The family of vectors (\ref{xncs})  resolves  the unity :
\begin{align}
\label{rengenquant}
\nonumber \int_{\mathcal{D}_{\sqrt{R}}}\nu(d^2z) \,  |v_z\rg\lg v_z| &= \sum_{n, n'= 0}^{\infty}  |e_n\rg  \lg e_{n'}|\, \frac{1}{\sqrt{x_n! x_n'!}}\, \int_{\mathcal{D}_{\sqrt{R}}}\dfrac{d^2 z}{\pi}\,  w(\vert z \vert^2)  z^n{\bar z}^{n'}\\[0.30cm]
 & = \sum_{n= 0}^{\infty}  |e_n\rg  \lg e_{n}|\, \frac{1}{x_n! }\, \int_{0}^R dt\,  w(t )\; t^n = I_d\, ,
\end{align}
with the last equality resulting trivially from (\ref{genmom}) and (\ref{scprod}).
\end{itemize}
Vectors  (\ref{xncs})  are known in the literature as ``non-linear'' coherent states, an attributive adjective  essentially issued from Quantum Optics \cite{matosvogel96,manko96,manko97,sanchez11} (see also \cite{dodonov02} for exhaustive references).

Property (v) is fundamental for quantization of the open disk $\mathcal{D}_{\sqrt{R}}$ in the sense that it allows one to define:

\begin{enumerate}
  \item a normalized positive operator-valued measure (POVM) on 
 $\mathcal{D}_{\sqrt{R}}$  equipped with the  measure $\nu(d^2z) $
  \eqref{modmeas} and its $\sigma-$algebra $\mathcal{F}$ of Borel sets :
\begin{equation*}
\label{stpov}
\mathcal{F} \ni \Delta \, \mapsto \int_{\Delta}\nu(d^2z) \,  |v_z\rg\lg v_z| \,
\, \in \mathcal{L}(\mathcal{H})^+\, ,
\end{equation*}
where $\mathcal{L}(\mathcal{H})^+$ is the cone of positive bounded
operators on $\H$;
  \item a  Berezin-Klauder-Toeplitz (or ``anti-Wick'') quantization,  simply named \emph{coherent state quantization throughout this paper}, of the complex
  plane \cite{klauder63,berezin75,klauder95,gazeau09}, which means that to a function
  $f(z,\bar{z})$ in the complex plane  corresponds the operator
  $A_f$ in $\H$ defined by
  \begin{equation}
\label{stquant}
f \mapsto A_f = \int_{\mathcal{D}_{\sqrt{R}}}\nu(d^2z) \, f(z,\bar{z}) \, |v_z\rg\lg v_z| =
\sum_{n, n'= 0}^{\infty} \left(A_f\right)_{nn'} \, |e_n\rg  \lg e_{n'}|\,,
\end{equation}
with matrix elements
\begin{equation*}
\label{matel1}
\left(A_f\right)_{nn'}= \frac{1}{\sqrt{x_n!x_{n'}!}}\, \int_{\mathcal{D}_{\sqrt{R}}}\frac{d^2z}{\pi}\, w(\vert z \vert^2)\, 
f(z,\bar{z}) \, 
z^n{\bar z}^{n'}\, .
\end{equation*}
The operator-valued integral (\ref{stquant}) is  understood  as the sesquilinear form, 
\begin{equation*}
B_f(\psi_1,\psi_2)= \int_{\mathcal{D}_{\sqrt{R}}}\nu(d^2z) \, f(z,\bar{z}) 
\,\lg \psi_1|v_z\rg\lg v_z|\psi_2\rg.
\end{equation*}
The form $B_f$ is  assumed to be  defined on a dense subspace of the Hilbert space.  If $f$ is real and at least semi-bounded, the Friedrich's extension of $B_f$ univocally defines a self-adjoint operator. However, if $f$ is not semi-bounded, there is no natural choice of a self-adjoint operator associated with $B_f$. In this case, we can consider directly the symmetric operator $A_f$  enabling us to obtain a self-adjoint extension (unique for particular operators). The question of what is the class of operators that may be so represented is a subtle one. We will come back to this point in Definition \ref{defobscs} below.

 If the function $f$ depends on $\vert z \vert^2 =t$ only, we get (formally) a diagonal operator with matrix elements
 \begin{equation*}
\label{mateldiag1}
\left(A_f\right)_{nn'}= \frac{\delta_{n n'}}{x_n!}\, \int_0^R dt \, w(t)\, t^n \, f(t) \, .
\end{equation*}
On the other hand, if $f$ depends on the angle $\theta = \arg z$ only, i.e. $f(z,\bar z) = F(\theta)$, the matrix elements of $A_f$ involve the Fourier coefficients  
$$
c_n(F) = \displaystyle \frac{1}{2\pi}\int_0^{2\pi}d\theta\, e^{-in\theta}\, F(\theta) ,
$$
i.e.
 \begin{equation*}
\label{matelang1}
\left(A_F\right)_{nn'}= c_{n'-n}(F)\, \frac{x_{\frac{n+n'}{2}}!}{\sqrt{x_n!\;x_{n'}!}} \, .
\end{equation*}
For instance, the following self-adjoint ``angle'' operator is defined in this way. Its matrix elements are given by:
 \begin{equation}
\label{angleop1}
A_{\theta}= \pi\,  I_{{\mathcal H}} + i \, \sum_{n\neq n'}\frac{x_{\frac{n+n'}{2}}!}{\sqrt{x_n!\;x_{n'}!}}\, \frac{1}{n'-n}\, |e_n\rg\lg e_{n'}|\, ,
\end{equation}
where we recall that  $x_{\nu}!$ is defined by   $x_{\nu}! = \int_0^R dt\,w(t)\,t^{\nu}$ whenever the latter integral converges.
\end{enumerate}

For the elementary functions $f(z,\bar{z}) = z$ and $f(z,\bar{z}) = \bar{z}$ we obtain
lowering and raising operators, respectively,
\begin{align}
   A_z &= a\, , \; \qquad a \, | e_n \rg = \sqrt{x_n} | e_{n-1}\rg\, ,
   \quad\; a\,| e_0\rg  = 0 \,  \quad \mbox{(lowering operator)} \label{stoper1},\\[0.30cm]
    A_{\bar z} & = a^{\dag} \, , \qquad a^{\dag} \, | e_n \rg =
     \sqrt{x_{n+1}} |e_{n+1}\rg \qquad \qquad\qquad \mbox{(raising operator)}\, . 
     \label{stoper2}
\end{align}
It results that the state $|v_z\rg$ is eigenvector of $A_z = a$ with eigenvalue $z$ (i.e. 
$a |v_z\rg = z|v_z\rg$) like for standard CS. 
Operators $a$ and $a^{\dag}$ obey the generically non-canonical commutation rule 
$$
[a,a^{\dag}]= x_{N+1} - x_{N}, 
$$ where the ``deformed-number'' operator $x_{N}$ is defined by
$x_{N} = a^{\dag} a$ and is such that its spectrum is
exactly $\{x_n\, , \, n \in \N\}$ with eigenvectors $x_{N}\,| e_n\rg=
x_n |e_n \rg $. The linear span of the triple $\{a, a^{\dag},
x_N\} $  is obviously not closed, in general,  under commutation and the set
of resulting commutators  generically gives  rise to an infinite
dimensional Lie algebra. These algebraic structures were extensively  examined in previous papers (e.g. \cite{arik92}).  

The next simplest function on the complex plane is 
$f(z,\bar{z}) = \vert z \vert^2$. 
When we consider the complex plane as the phase space of a
particle moving on the line, i.e. $z = \frac{1}{\sqrt{2}}(\q + i\p)$ and, hence, 
$\vert z \vert^2 = \frac{1}{2}(\p^2 + \q^2)$ we get, in appropriate units,  the classical Hamiltonian of the harmonic oscillator (H.O.). In this case 
\begin{equation*}
\label{qcsho}
A_{z\bar z} = A_z\, A_{\bar z}= a\, a^{\dag} = x_{N+1}\, .
\end{equation*}
Therefore, the spectrum of the quantized version of $\vert z \vert^2 $  is the 
 sequence $\left( x_n\right)_{n\geq 1}$. 
 
 Let us consider the (deformed) position and momentum operators as the respective CS quantizations of the phase space coordinates $\q$ and $\p$:
\begin{equation}\label{QP-operators}
 A_\q \equiv Q = \frac 1{\sqrt{2}}(a + a^\dagger)\; , \qquad
 A_\p \equiv P = \frac 1{i\sqrt{2}}(a - a^\dagger)\; .
\end{equation}
 It is worthy to note that if we had adopted the usual procedure of the canonical quantization for defining the quantum version of a classical observable $f(\q,\p)$, namely replacing $\q$ and 
 $\p$ by $Q$ and $P$ respectively in the latter and proceeding with the symmetrization of $f(Q,P)$, we would have obtained expressions in general different of $A_f$. In the case of simple quadratic expressions these discrepancies read as  
 \begin{equation}
\label{q2p2}
A_{q^2}= Q^2 + \frac{1}{2}(x_{N+1}-x_N)\,, \quad A_{p^2}= P^2 + \frac{1}{2}(x_{N+1}-x_N)\, ,  
\end{equation}
 and so for the H.O. Hamiltonian, 
  \begin{equation}
\label{q2+p2}
\frac{1}{2}(P^2 + Q^2) = A_{(p^2+q^2)/2} -\frac{1}{2}(x_{N+1}-x_N) = \frac{1}{2}(x_{N+1}+x_N)\, .   
\end{equation}
Now an energy is always defined up to a constant. One usually decides that the zero-point of quantum energies lies at the infimum  of the spectrum ($\sigma$) of the quantum potential energy. In the case of the canonical quantization, the latter  is $\inf\sigma(Q^2/2)$ whilst  it is $\inf\sigma((Q^2 + x_{N+1}-x_N)/2)$ in the case of  CS quantization. So we have to compare the spectral values $x_n -  \inf\sigma(Q^2/2)$ (for the canonical quant.) with $x_{n+1} - \inf\sigma((Q^2 + x_{N+1}-x_N)/2)$ (for the CS quant.). To solve this problem, we need to know more about the sequence $(x_n)$. In the non-negative integer case, the two above differences are identical. 


Given a function $f$ on the  complex plane, the resulting operator 
$A_f$,
if it exists, at least  in a weak sense, acts on the Hilbert space ${\mathcal H}$. The  integral
\begin{equation*}
\label{wksssymb}
\lg\psi | A_f |\psi\rg= \int_{\mathcal{D}_{\sqrt{R}}}f(z,\bar{z})\vert \lg\psi |v_z\rg\vert^2  \,
\nu(d^2z)
\end{equation*}
should be finite for all  $\psi$ in some dense subset  of
$\mathcal{H}$. 
In order to be more rigorous on this important point, let us adopt the following
acceptance criteria for a function  to belong to the class of quantizable classical
observables.
\begin{defi}
\label{defobscs}
A  function $\mathcal{D}_{\sqrt{R}} \ni z \mapsto f(z,\bar{z}) \in \C$  is a \emph{CS quantizable
classical observable} via the  map $f \mapsto A_f$ defined by
(\ref{stquant})
if the map $\mathcal{D}_{\sqrt{R}} \ni z = \frac{1}{2} (\q + i\p) \equiv (\q,\p) \mapsto \lg v_z |
A_f |v_z \rg$  is a smooth (i.e. $ C^{\infty}$) function
with respect to the $(\q,\p)$ coordinates.
 \end{defi}

 The function $f$  is the \emph{upper} \cite{csfks} or
  \emph{contravariant}
 \cite{berezin75} symbol of the operator $A_f$, and the mean value of the latter in state $ |v_z \rg$, 
 \begin{equation*}
\label{lowsymb1}
\check{f}(z,\bar z)\deq   \lg v_z |A_f |v_z \rg = \int_{\mathcal{D}_{\sqrt{R}}}f(z',\bar{z'})\vert \lg v_z |v_{z'}\rg\vert^2  \, \nu(d^2z')\, , 
\end{equation*}
 is the \emph{lower} \cite{csfks} or \emph{covariant}
 \cite{berezin75} symbol of the operator $A_f$. The map $f \mapsto \check{f}$ is an integral transform with kernel  $\vert \lg v_z |v_{z'}\rg\vert^2 $ which generalizes the Berezin transform.

In \cite{ChaGaY08} the  definition \ref{defobscs}  in the standard case $x_n = n$ is extended to a class of  distributions including tempered distributions.

Localization properties in the complex plane, from the point of view of the
sequence $\{ x_n \}_{n\in \N}$, 
should be examined through the shape (as functions of  $z$) of the respective
lower symbols of $Q$ and $P$:
\begin{equation*}
\label{lowsymbQP}
\check{Q}(z) \deq \lg v_z |Q | v_z\rg\, ,
\qquad \check{P}(z) \deq \lg v_z |P | v_z\rg\, ,
\end{equation*}
and the noncommutativity  reading of the complex plane should be
encoded in the behaviour of the lower symbol  $\lg v_z |[ Q,
P] | v_z\rg$ of the commutator $[Q, P]$.
The dispersions verify
\begin{equation}
\label{uncert}
\left(\Delta_{v_z}Q\right)^2= \left(\Delta_{v_z} P\right)^2 = \frac{1}{2}\, \lg v_z |(x_{N+1} - x_N)| v_z\rg  
=\frac 12 \vert  \lg v_z |[Q,P]| v_z\rg \vert 
\end{equation}
for any $| v_z\rg $, which indicates that the latter form a family of \emph{intelligent} states. 
The study, within the above framework, of the product of dispersions expressed in states $|v_z\rg $,
\begin{equation}
\label{uncert1}
 \left(\Delta_{v_z}Q\right)\, \left(\Delta_{v_z} P\right) = \frac{1}{2}\,
  \lg v_z |(x_{N+1} - x_N)| v_z\rg  \, , 
\end{equation}
 should thus be relevant since  they saturate the uncertainty relations within the quantization context provided by them.  Note that eq. (\ref{uncert1}) can also be written in terms of mean values $\lg F(n)\rg$ of  discrete function $n \mapsto F(n)$ with respect to the Poisson-like distribution $n \mapsto \dfrac{t^n}{\mathcal{N}(t)\;x_n!}$:
\begin{equation*}
\label{uncert2}
 \left(\Delta_{v_z}Q\right)\, \left(\Delta_{v_z} P\right) 
=  \frac{1}{2}\, \langle  (x_{n+1} - x_n)\rangle = \frac{1}{2}\, \left(\langle x_{n+1}\rg - \vert z \vert^2\right)\, . 
\end{equation*}
Another interesting function emerging from this formalism is the lower symbol of a function $F(\theta)$ of the angle only. With $z= re^{i\theta}$ we find the series
\begin{equation}
\label{fctangleop2}
 \check{F}(z,\bar z)=\lg v_z |A_F |v_z\rg
=  c_0(F) +\sum_{k\neq 0} d_{|k|}(r)\,c_k(F)\,e^{ik\theta}\, , 
\end{equation}
where the function 
\begin{equation}
\label{dkr}
d_{|k|}(r)= \frac{r^k}{\mathcal{N}(r^2)}\sum_{n=0}^{\infty} \frac{x_{\frac{|k|}{2}+n}!}{x_n!\; x_{n+|k|}!}\,r^{2n}
\end{equation}
balances the Fourier coefficient $c_k(F)$ of the function $F$ (see Appendix A).

Thus for  the angle operator $A_\theta$ (\ref{angleop1}) we have the particular series (see Fig.~\ref{low_symbol-angle1}a):
\begin{equation}
\label{angleop2}
 \check{\theta}(z,\bar z) =  \pi+ i\sum_{k\in \Z-\{0\}} \frac{d_{|k|}(r)}{k}\,e^{ik\theta}\, =
  \pi-2 \sum_{k=1} ^{\infty} \frac{d_{|k|}(r)}{k}\,\sin ({k\theta})\, . 
\end{equation}
Note that if we wish to restore physical units, e.g. for the motion of a particle (mass $m$) on the line, we should introduce some characteristic length  and momentum which will provide unit standard for the $(\q,\p)$ variables, say $\ell_c$ and $\wp_c$ respectively. Since we have in view a (CS) quantum version of this motion, it is natural to impose the relation $\ell_c\, \wp_c= \hbar$.  So we keep dimensionless the phase space variable $z$ by putting
\begin{equation}
\label{dimphsp}
z = \frac{1}{\sqrt{2}}\left(\frac{\q}{\ell_c} + i \ell_c\frac{\p}{\hbar}\right)\, . 
\end{equation}
It is clear from the above expression that the (semi-)classical regime corresponds to large $\vert z \vert  =r$ and that the original function $F(\theta)$ is recovered through its Fourier series if $\lim_{r\to \infty} d_k(r) = 1$  for each $k$. 
 
Let us tell more about the physical meaning of such a coherent state quantization, although if we go back to dimensionless quantities for the sake of simplicity. In the case of the sequence of non-negative integers, we have $x_n = n$, $R = \infty$ and we recover the canonical quantization of the complex plane $\mathcal{D}_{\infty} = \C$  viewed as the classical phase space for the motion of a particle on the line, equipped with the usual Lebesgue (or ``uniform'') measure $\nu(d^2z)=  d^2z/\pi$. Within this standard scheme, the quantized version of the classical H.O. Hamiltonian $H = \frac{1}{2} (\p^2 + \q^2)$ is the number operator $N+1$ with spectrum $1,2,\dotsc$. Also, one can prove in this case that  
$\lim_{r\to \infty} d_k(r) = 1$ for each $k$  (see Fig.~\ref{fig_drk}a). 
Now, changing the measure $d^2z/\pi$ into  
\begin{equation*}
\nu(d^2z) = w(\vert z \vert^2) \,\mathcal{N} (\vert z \vert^2) \,  
\mathrm{H}(\vert z \vert^2 -R) \, d^2z/\pi ,
\end{equation*} 
where $\mathrm{H}(\cdot)$ is the Heaviside function, we give to  the phase space a different statistical content: classical states are not anymore described by points $z_0$, i.e. by Dirac distributions $z \mapsto \delta_{z_0} (z)$, but instead by the distribution  
$ z \mapsto w(\vert z -z_0 \vert^2)\mathcal{N}(\vert z -z_0\vert^2)\,  \mathrm{H}(\vert z \vert^2 -R) $. Then, by following the
CS quantization stemming from the associated sequence $\left(x_n \right)_{n\in\N}$, we consistently find the latter as another (possibly observed!) energy spectrum along the equation
\begin{equation*}
\label{qcsho1}
A_{\frac{1}{2} (p^2 + q^2)}= A_{z\bar z} = a\, a^{\dag} = x_{N+1}\, .
\end{equation*}

Since we are dealing with a different  quantized version of
the classical observable $|z|^{2}$ and considering it as a quantum Hamiltonian $\hat{H}$, ruling the time
evolution of quantum states, it is natural to  investigate the time evolution of the
quantized version $a = A_z$, as found in (\ref{stoper1}), of the
classical phase space point $z= (\q+i\p)/\sqrt{2}$, comparing it with
the phase space circular classical trajectories. This time evolution is well
caught through its mean value in coherent states $|v_z\rangle$ (lower
symbol) :
\begin{equation}\label{timeev} 
\begin{array}{lll}
  \check{z}\left(  t\right)  &\overset{\mathrm{def}}{=}&\left\langle
v_z\right\vert e^{-i\hat{H}t}\,  A_z \, e^{i\hat{H}t}\left\vert
v_z\right\rangle  \\[0.3cm]
 &=&\displaystyle
\frac{z}{\mathcal{N}_{\lambda}(|z|^{2})}\sum_{n=0}^{+\infty} 
\frac{\left\vert z\right\vert ^{2n}}{x_{n}!}\,\exp\left(  {i}\left(
{x_{n+2}-x_{n+1}}\right)  {t}\right)  \,.
\end{array}
\end{equation}
In the standard case we just have $  \check{z}\left(  t\right) = e^{it}z$, which describes nothing else but the classical phase-space trajectory. In the deformed case (\ref{timeev}) we easily prove that $ \vert \check{z}\left(  t\right)\vert \leq \vert z \vert$. 

There exists another interesting exploration of the semi-classical character of coherent states. For this purpose  we can define a phase space distribution in terms of the following density of probability (given a fixed a normalized state, say $|v_{z_0}\rangle$) on the complex plane equipped with the measure $d\nu(=d^2 z)$ for all $t$ :
\begin{equation}
\label{probdensphs}
z=(q,p)/\sqrt{2}\in \C \mapsto \rho_{v_{z_0}}(z) \deq \vert \lg v_z|v_{z_0}\rg \vert^2= \frac{\vert \mathcal{N}(\bar z z_0 )\vert^2}{\mathcal{N}(\vert z\vert^2)\mathcal{N}(\vert z_0\vert^2)}\, .
\end{equation}
The time evolution behavior of the probability density $ \rho_{v_{z_0}}(z)$  is given by
\begin{equation}
\label{probdens}
z \mapsto \rho_{v_{z_0}}(z,t) \deq \vert \lg v_z| e^{-i\hat Ht}|v_{z_0}\rg \vert^2= \frac{\vert \mathcal{N}_{\Delta_N\,t}(\bar z z_0 )\vert^2}{\mathcal{N}(\vert z\vert^2)\mathcal{N}(\vert z_0\vert^2)}\, ,
\end{equation}
where the modified ``exponential'' appearing at the numerator is defined by:
\begin{equation*}
\label{modexp}
\mathcal{N}_{\Delta_N\,t}(z ) \deq \sum_{n=0}^{\infty}\frac{z^n}{x_n!\; e^{i\Delta_n t}}\, , \quad \Delta_n= x_{n+1} \, . 
\end{equation*}


\section{Integers as deformations of integers}
\label{Integersasdeformationsofintegers}

It is well-known that in the $q$-deformation of Lie algebras and in the study of their $q$-representations appear  some real numbers that are deformations of nonnegative  integer numbers like
\begin{equation}\label{qns}
[n]_q:=\frac{1-q^{\pm n}}{1-q^{\pm 1}}, \qquad  {}^s[n]_q:=\frac{q^{ n}-q^{- n}}{q-q^{-1}} ,  \quad n\in \N,\; q\in \R,
\end{equation}
since in the limit $q\to 1$ we recover the original number $n$.
For  the $q$-oscillator  these $q$-integer numbers constituted the energy spectrum of deformed versions of the harmonic oscillator ($0<|q|<1$).
Moreover, $q$-coherent states are constructed for these  $q$-oscillators.
 
In general, these sequences of $q$-numbers   
are non-integer numbers. Hence, one is naturally led to address the question:
do there exist such  sequences with $[n]_q\in \N\;\; \forall n\;\in \N$?
For the previous $q$-deformation $[n]_q$ the answer is trivially no  since if
\[
[n]_q\equiv\frac{1-q^n}{1-q}=1 + q + q^2+\cdots + q^{n-1}\quad \mbox{for any} \, n
\]
then $q\in \N$.

 However, there are other deformations of numbers where  it is possible  to obtain  sequences  of non-negative integers.    The interest of such sequences would be to get, for instance,  $qp$-harmonic oscillator with  spectrum of positive integer numbers.
As a matter of fact, let us consider the so-called $pq$-deformations of non-negative integers (or natural numbers) $n\in \N$: 
\begin{equation*}
\label{qpdef}
[[n]]_{qp} := \frac{q^n - p^n}{q-p}\, .
\end{equation*}
The particular cases $p = 1$ and $p = q^{-1}$ correspond to $q$-deformations \eqref{qns}.

Now, we demand  that the following property holds for  
{any} $n \in \N$:
\begin{equation*}
\label{condinteg}
u_n \equiv [[n]]_{qp} \in \N\, . 
\end{equation*}
We already check that it is true for $u_0 = 0$ and $u_1 = 1$. From $q + p = u_2 \in \N$ and $q^2 + qp + p^2 = u_3 \in N$, we immediately infer that 
$q$ and $p$ are quadratic integers, i.e. both are  roots of the quadratic equation
\begin{equation}
\label{qpequat}
X^2 -sX + r = 0\, , \quad s = u_2 = p+q \in \N \, , \quad r = pq = u_2^2 - u_3 \in \N\, .
\end{equation}
We now impose that   both $q$ and $p$ be non-zero real numbers. This implies that $ 0 \leq s^2 -4r = 4 u_3 - 3 u_2^2 = u_3 -3 r$. So, from $u_3\in \N$ and $r \in  \Z$ we infer that $q$ and $p$ are real if $r< 0$, i.e. $r$ is a negative integer. If $r>0$, i.e. $r$ is a positive integer, then $u_3$ should be bigger than $3 r$. Hence, the choice of $r, s\in \Z^{\ast}$ determines completely the sequence of the $u_n$'s through a three-term recurrence which replicates exactly the algebraic equation (\ref{qpequat}). Indeed, by definition we have $u_0 = 0$, $u_1 = 1$ and $u_2 = p+q$. Next,  
\[
u_3 = (p+q)^2 -qp = u^2_2 - r = su_2 -r ,
\] 
and more generally, from the trivial identity, 
\begin{equation*}
p^{n+1}- q^{n+1}= (p+q)(p^n - q^n) + pq (p^{n-1}- q^{n-1})\, 
\end{equation*}
 we have
\begin{equation}
\label{unequat}
u_{n+1} -su_n + r u_{n-1} = 0\, , \quad  u_0 = 0\, , \quad u_1 = 1\,  .
\end{equation}
 Such sequences of numbers generalize the famous Fibonacci sequence corresponding to the simplest case $s = 1$ and $r=-1$. In this case, $p = \dfrac{1 + \sqrt{5}}{2} \equiv \tau$ (the golden mean) and $q = \dfrac{1 - \sqrt{5}}{2}= - 1 /\tau$. 
 
 In this paper, we precisely focus on the cases $r= \pm 1$ and $s> 0$  (the  choice $s<0$  corresponding to a change of sign for both roots).

\begin{itemize}
\item[ ]\textbf{Case 1.-}
 If we choose $r= -1$, then $s\geq 1$, and the roots of 
 \begin{equation*}
\label{qpequatunit-}
X^2 -sX -1 = 0\,   
\end{equation*}
are $q,p$ such that, say $-1<q<0$ and $1 <p $.
 
\item[ ]\textbf{Case 2.-}
 If we choose instead $r = 1$, then $ s\geq 2$ and the equation now reads 
 \begin{equation}
\label{qpequatunit+}
X^2 -sX +1 = 0\, . 
\end{equation}
We exclude the trivial case $s =  2$  which corresponds to $p= q = 1$, and we will consider all other cases, $s\geq 3$,  which give $0<q<1$ and $p = 1/q >1$. 
 \end{itemize}

 In both cases, the  algebraic integer $p> 1$ is called a {\bf quadratic} (because of the degree of the equation)  {\bf Pisot-Vijayaraghavan} (because $p> 1$ the other root, $q$ has modulus less than 1) {\bf unit } (since $r = \pm 1$). The complete description of the cases of quadratic Pisot numbers which are  not unit is given in the appendices (see \cite{boyd} and references therein).

In case 2, we are precisely in the situation of the so-called symmetric or bosonic $q$-deformation of natural numbers:
\begin{equation}
\label{qsdef}
{}^{[s]}[n]_{q} = \frac{q^n - q^{-n}}{q-q^{-1}} = {}^{[s]}[n]_{q^{-1}}\, .
\end{equation}

In case 1, we are instead in the situation of antisymmetric or fermionic $q$-deformation of natural numbers:
\begin{equation*}
\label{qsdeff}
{}^{[f]}[n]_{q} =  \frac{q^n - (-1)^nq^{-n}}{q +q^{-1}}= (-1)^{n}\, {}^{[f]}[n]_{q^{-1}}\, .
\end{equation*}

\begin{center}
	\begin{tabular}{|l||c|c|c|c|c|c|c|c|c|c|c|c|c|} \hline
	 \multicolumn{1}{|c||}{\bf Numbers}& \multicolumn{1}{|c|}{\bf n=1} & \multicolumn{1}{c|}{\bf 2} &\multicolumn{1}{c|}{\bf 3} &\multicolumn{1}{c|}{\bf 4}& \multicolumn{1}{c|}{\bf 5} &\multicolumn{1}{c|}{\bf 6} &\multicolumn{1}{c|}{\bf 7} &\multicolumn{1}{c|}{\bf 8} &\multicolumn{1}{c|}{\bf 9} &\multicolumn{1}{c|}{\bf 10} &\multicolumn{1}{c|}{\bf 11} \\  \hline \hline
 {\bf Fibonacci}& 1 &2 &3&5&8&13&21&34&55&89&144\\  \hline
{\bf q-fermionic}& 1 &2 & 5 &12&29&70&169&408&985&2378&5741\\  \hline
{\bf q-bosonic 1} & 1 &3 &8&21&55&144&377&987&2584&6765&17711 \\  \hline
{\bf q-bosonic 2} & 1 &4 &15&56&209&780&2911&10864&40545&151316&564719 \\  \hline
{\bf q-bosonic 3} & 1 &5 &24&115&551&2640&12649&60605&290376&1391275&6665999
\\  \hline
\end{tabular}\medskip

  {\footnotesize Table 1: Quadratic Pisot numbers with $q=(1-\sqrt{5})/2,\;q= 1-\sqrt{2},\;q= (3-\sqrt{5})/2,\;q= 2-\sqrt{3},\; q= (5 - \sqrt{21})/2. $}\label{table_pisot}\end{center}
\vskip-0.5cm

\begin{figure}[h]
\centering
\includegraphics[scale=1]{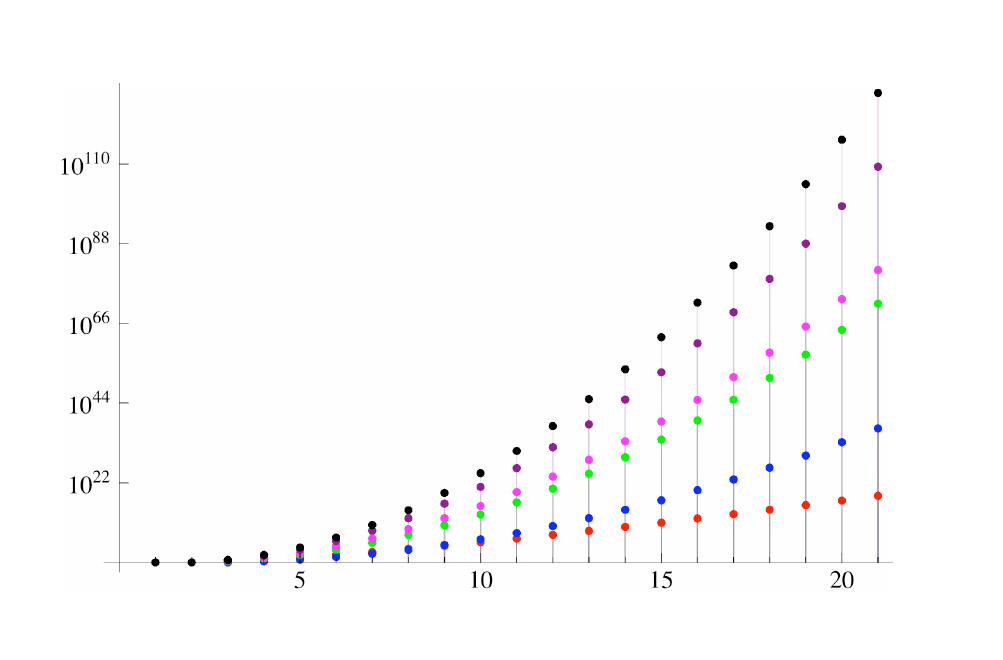}\vskip-0.75cm

\caption{\footnotesize Factorials of  Pisot numbers  together with the standard factorial (red) versus $n$ for   $q=(1-\sqrt{5})/2$ (blue),  $q= (1-\sqrt{2})$ (green), $q= (3-\sqrt{5})/2$ (magenta), $q= (2-\sqrt{3}$ (purple), $q= (5-\sqrt{21})/2$ (black).} \label{fig_factorials}
\end{figure}

\section{Moment measure for the symmetric deformation of integers}\label{momentumproblem}

To a given $q >0$  let us associate the sequence $\left( x_n\right)_{n\in \N}$ defined by the symmetric deformations of nonnegative integers $(x_n\equiv {}^{[s]}[n]_{q})_{n\in \N}$ given by \eqref{qsdef}. 
Our aim is to make explicit the moment measure for this sequence,  i.e. to  find a probability distribution $t \mapsto w_q(t)$  on $[0, R )$
such that
\begin{equation*}
\label{genmomentum}
x_n! = {}^{[s]}[n]_{q}!= \int_{0}^{R} t^n \, w_q(t)\, dt\, .
\end{equation*}

Since
\begin{equation*}
R = \lim_{n \to \infty} x_n = \infty\, , \quad \forall\ q > 0\, , 
\end{equation*}
its associated exponential 
\begin{equation}
\label{seqexp1}
\mathfrak{e}_q(t) \equiv  \mathcal{N}_q(t)= \sum_{n = 0}^{+\infty} \frac{t^n}{x_n!}
\end{equation}
defines an analytic entire function $\mathfrak{e}_q(z)$ in the complex plane for any positive $q$. This series coincides with  $\mathcal{N}(t)$ \eqref{seqexp} that now we call $\mathcal{N}_q(t)$ since $x_n$ depends on $q$.
\begin{figure}[h]
\centering
\includegraphics[scale=.95]{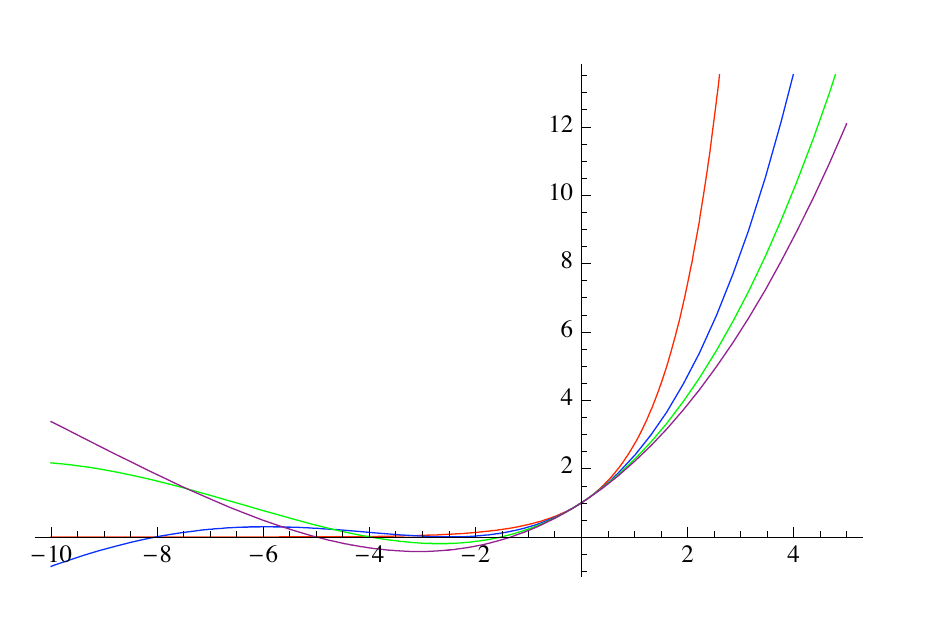}
\vskip-0.5cm

\caption{\footnotesize $\mathfrak{e}_q(t)$  for $q=1$ (red), $q=(3-\sqrt{5})/2$ (blue), $q= 2-\sqrt{3}$ (green), $q= (5-\sqrt{21})/2$ (purple).} \label{fig_exponentials}
\end{figure}

Let us now  introduce the ``auxiliary'' exponential :
\begin{equation*}
\label{seqexp2}
 \mathfrak{E}_q (t)   \deq \sum_{n=0}^{\infty} q^{\frac{n(n+1)}{2}}\frac{t^n}{x_n!} 
 \equiv\sum_{n=0}^{\infty} \frac{t^n}{y_n!}\, , 
\end{equation*}
where $y_n \deq q^{-n} x_n$.
Its radius of convergence is $\infty$ for $0< q \leq 1$ and is equal to $1/(q-q^{-1})$ for $q> 1$.
\begin{figure}[h]
\centering
\includegraphics[scale=.95]{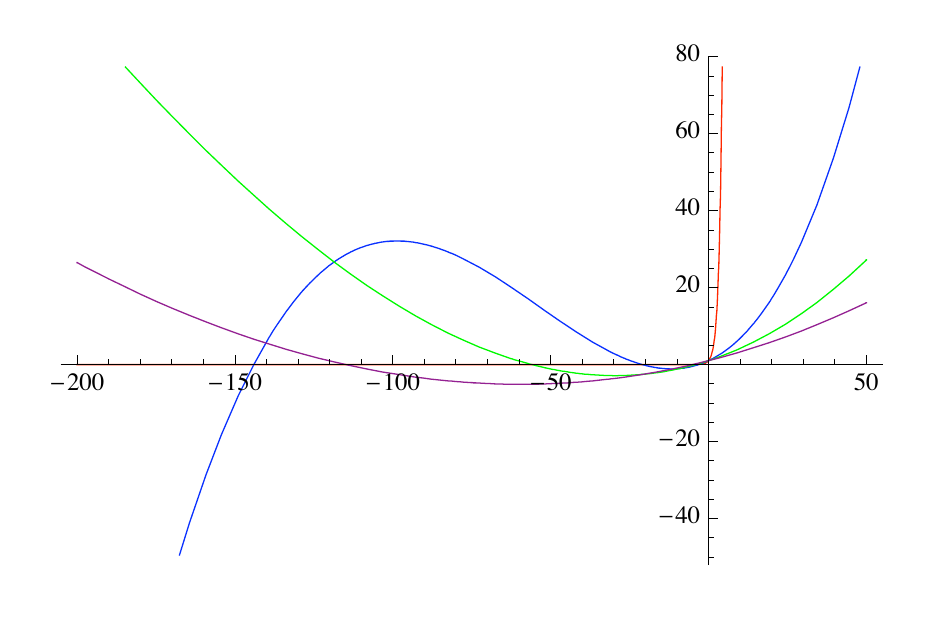}
\vskip-0.5cm

\caption{\footnotesize $\mathfrak{E}_q(t)$  for $q=1$ (red), $q=(3-\sqrt{5})/2$ (blue), $q= 2-\sqrt{3}$ (green), $q= (5-\sqrt{21})/2$ (purple).} \label{fig_mod-exponentials}
\end{figure}

This last exponential is connected with  the two standard $q$-exponentials as they are defined in \cite{solkac} (see Appendix B for more details):
\begin{equation}
\label{eEmfE}
\mathfrak{E}_q(t) = e^{qt}_{q^{-2}} = E^{qt}_{q^{2}}\, , 
\end{equation}
where
\begin{eqnarray}
\label{eqsol}
  e_q^x \deq   \sum_{n=0}^\infty \frac{x^n}{[n]_q!}\quad \left(=
\frac{1}{ \prod_{j=0}^{\infty}(1-q^j(1-q)x)}\quad \mbox{for} \quad q< 1\right) \, ,  \nonumber \\ 
\label{Eqsol}
  E_q^x \deq  \sum_{n=0}^\infty q^{n(n-1)/2}\frac{x^n}{[n]_q!}= e_{q^{-1}}^x \quad \left(= \prod_{j=0}^{\infty}(1+q^j(1-q)x)\quad \mbox{for} \quad q<1\right) \, , 
\end{eqnarray}
and  
\begin{equation*}
\label{defnq}
[n]_q \deq \frac{1- q^{n}}{1 - q}. 
\end{equation*}
We note that the series $e_q^x$ converges for all $x$ if $q \geq 1$ and for all $x$ such that $\vert x \vert < 1/(1-q)$ if $q<1$. On the  contrary,  $E_q^x$ converges for all $x$ if $q \leq 1$ and for all $x$ such that $\vert x \vert < 1/(1-q^{-1})$ if $q> 1$.
From the  relation $e_q^{-x} E_q^x = 1$, valid for all $x$ such that $\vert x \vert < 1/(1-q)$ if $q<1$ and  for all $x$ such that $\vert x \vert < 1/(1-q^{-1})$ if $q> 1$, we infer that, in the case $q\leq 1$, 
\begin{equation}
\label{invmfe}
\left\lbrack \mathfrak{E}_q(t)\right \rbrack^{-1} = \mathfrak{E}_{q^{-1}}(-t)\quad \forall\, t\quad \mbox{such that}\quad \vert t \vert < \frac{1}{q^{-1}-q}\, . 
\end{equation}
Also note that due to (\ref{eEmfE}) and (\ref{Eqsol}), the expression of  $\mathfrak{E}_q(t)$ as an infinite product for $q<1$ reads 
\begin{equation*}
\label{infprodEE}
 \mathfrak{E}_q(t) = \prod_{j=0}^{\infty}(1+q^{2j+1}\,(1-q^2)\,t)\quad \mbox{for}\quad \vert t \vert < \frac{1}{q(1-q^2)}\, , 
\end{equation*}
and so the first zero of $ \mathfrak{E}_q(t)$ standing on the left of the origin is equal to $-1/(q(1-q^2))$.
Due to these relations, the auxiliary exponential $\mathfrak{E}_q(t)$ is involved in  the $q$-integral representation(s) of the $q$-gamma function $\Gamma_q(x)$ defined \cite{solkac} by 
\begin{equation*}\label{defGq}
\G_q(x)=\frac{(1-q)_q^{x-1}}{(1-q)^{x-1}}\, ,\quad x>0\, , 
\end{equation*}
where the expression $(1+a)^x_q$ is given  in Appendix B. 
It obeys the functional equation
\begin{equation*}
\label{fcteqG}
\G_q(x + 1) = [x]_q\, \G_q(x)\, , \quad \mbox{for all}\ x>0\, , \quad \G_q( 1) = 1\, ,
\end{equation*}
and so
\begin{equation}
\label{fcteq2G}
\G_{q^2}(x + 1) = q^{x-1} \, {}^{[s]}[x]_q\, \G_{q^2}(x)\, .
\end{equation}
\begin{figure}[t]
\centering
 \subfigure [$\G_q(x)$ \hskip6.5cm (b) $\G_{q^2}(x)$]{
 \includegraphics[width=0.5\textwidth]{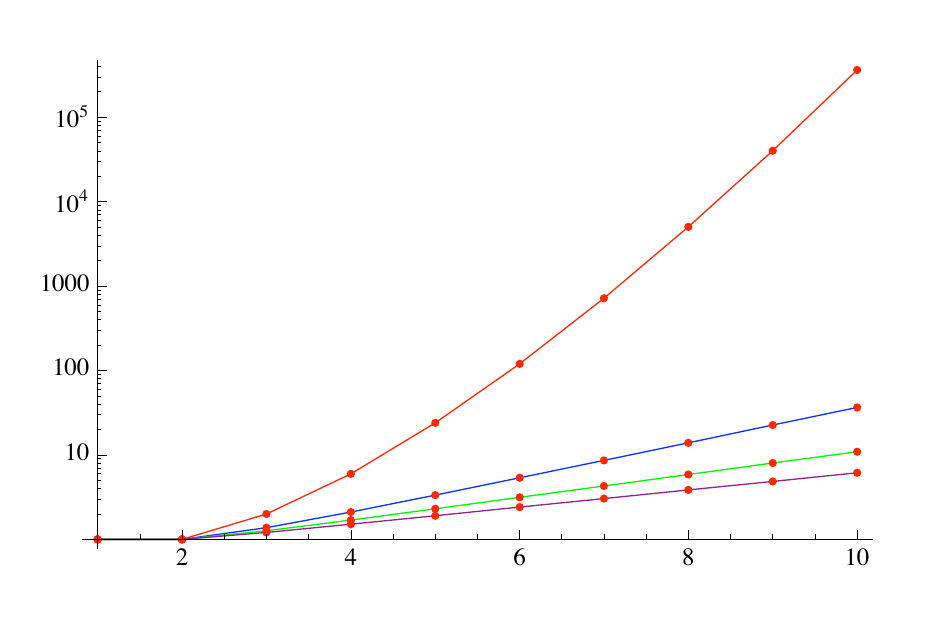}\qquad  
\includegraphics[width=0.5\textwidth]{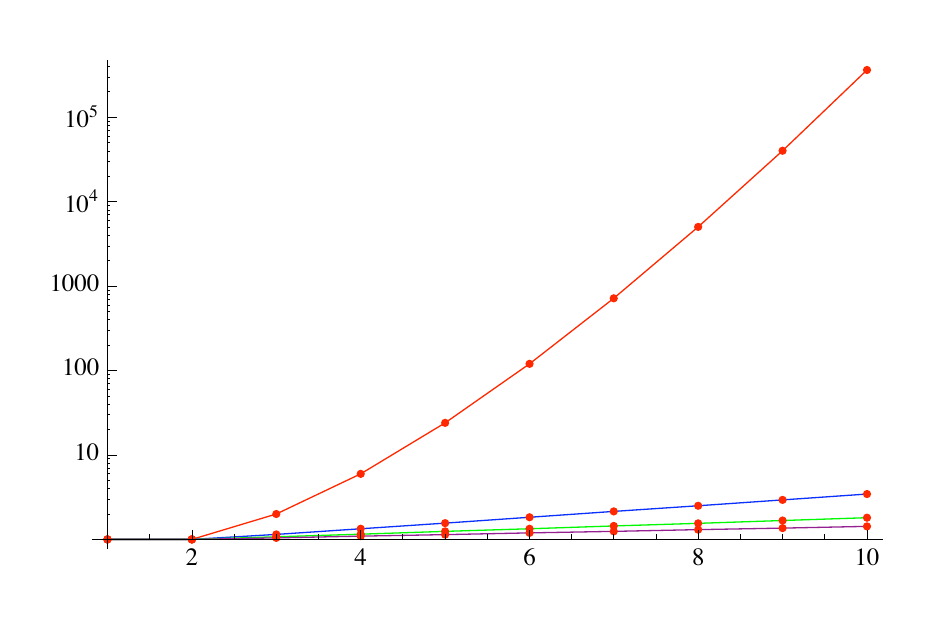}
}
\caption{\footnotesize Representations of $q$-gamma functions for $q=1$ (red), $q=(3-\sqrt{5})/2$ (blue), $q= 2-\sqrt{3}$ (green), $q= (5-\sqrt{21})/2$ (purple).}
\label{q-gamma}
\end{figure}

A first integral representation   is the following:
\begin{equation*}\label{intGq}
\G_q(x)=\int_0^{\frac{1}{1-q}}t^{x-1}E_q^{-qt}\, d_qt\ .
\end{equation*}
The $q$--integral (introduced by Thomae \cite{thomae} 
and Jackson \cite{jackson_1910}) is defined by
\begin{equation}\label{jackint}
\int_0^{a}f(t)\; d_qt
=(1-q)\sum_{j=0}^\infty a\;q^j\, f(a\;q^j)\, ,
\end{equation}
and can be  considered also as an ordinary integral with a discrete measure: 
\begin{equation}
\label{qdirmeas}
\int_0^{a}f(t)\; d_qt
= \int_0^{\infty} \rho_q(t, a) f(t)\, dt\, , \qquad 
\rho_q(t, a) = (1-q)\sum_{j=0}^\infty aq^j\, \delta(t-aq^j)\, .
 \end{equation}
Therefore, one derives from (\ref{jackint})  and (\ref{qdirmeas}) the integral representation
\begin{equation}
\label{intrepG2}
\G_q(x) = \int_0^{\infty} \rho_q(t, 1/(1-q)) \, t^{x-1}E_q^{-qt}\, dt\, .
\end{equation} 
Now, particularizing eq.\,(\ref{fcteq2G}) to nonnegative  integer values $x = n$ we obtain the expression
\begin{equation}
\label{Gnq}
\G_{q^2}(n + 1) = q^{\frac{n(n-1)}{2}}\, \qn\, \G_{q^2}(n) = q^{-n} \, q^{\frac{n(n+1)}{2}}\, x_n!\, .
\end{equation}
Combining \eqref{Gnq} with (\ref{intrepG2}), (\ref{eEmfE}), and the integration variable change $t \mapsto q\;t$,  we get the solution 
to the moment problem with a positive measure for the sequence 
$\left(q^{\frac{n(n+1)}{2}}\, x_n!\right)_{n\in \N}$ when $q < 1$:
\begin{equation}
\label{momeas}
 q^{\frac{n(n+1)}{2}}\, x_n! = \int_0^{\infty} t^{n}\,  \varpi_{q}(t) \,dt\, ,
\end{equation}
where $ \varpi_{q}(t)$ is the discrete measure:
\begin{equation*}
\label{varpiq}
 \varpi_{q}(t) = \sum_{j=0}^\infty q^{2j}\, \, \mathfrak{E}_q\left(-\frac{q^{2j}}{q^{-1}-q}\right)\,\delta\left(t- \frac{q^{2j}}{q^{-1}-q}\right)\,. 
\end{equation*}
An important point to be noted concerning this measure is its positiveness. Indeed, from (\ref{invmfe}) the factors  $\mathfrak{E}_q\left(-\frac{q^{2j}}{q^{-1}-q}\right)$ are positive for all $j\geq 0$. 

In order to solve the moment problem for the sole sequence $(x_n!)_{n\in \N}$ it is necessary to solve it for the sequence 
$\left( q^{-\frac{n(n+1)}{2}}\right)_{n\in \N}$ and to use an adapted composition formula \cite{berg1} for moments: let 
$a(t)$ and $b(t)$ be two weight functions solving the moment problems
\begin{equation*}
a_n! = \int_0^{\infty} t^n a(t)\, dt\, , \qquad 
b_n! = \int_0^{\infty} t^n b(t)\, dt\, , 
\end{equation*}
respectively. Then the weight function defined by the (multiplicative) convolution 
\begin{equation}
\label{convmel}
w(t) \deq   \int_0^{\infty}  a(t/u)\, b(u)\, \frac{du}{u}
\end{equation}
solves the moment problem 
\begin{equation*}
\label{convmelsol}
a_n!\, b_n! = \int_0^{\infty} t^n w(t)\, dt\, .
\end{equation*}
\begin{figure}[t]
\centering
 \subfigure {
 \includegraphics[width=0.485\textwidth]{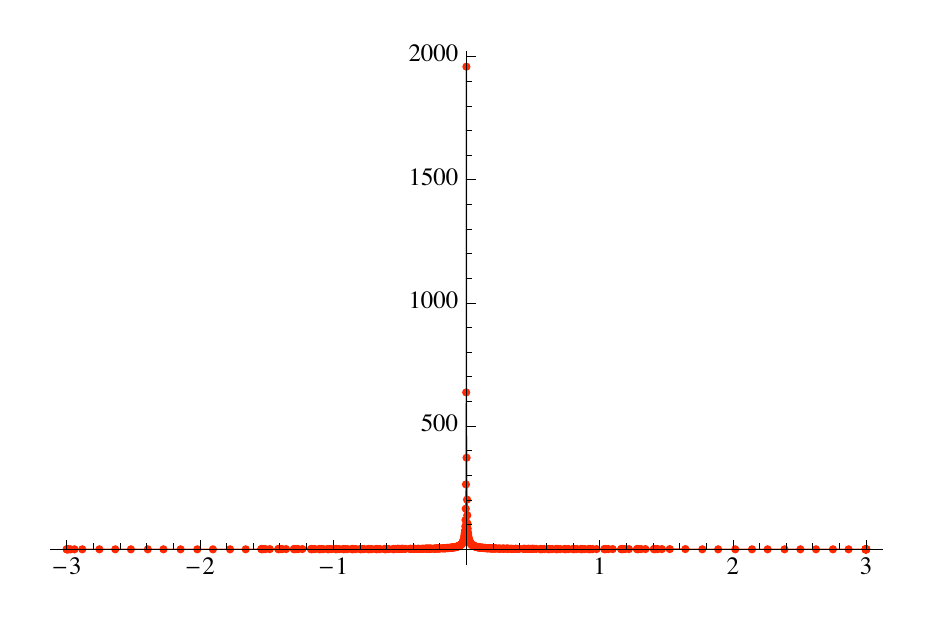}\qquad  
\includegraphics[width=0.485\textwidth]{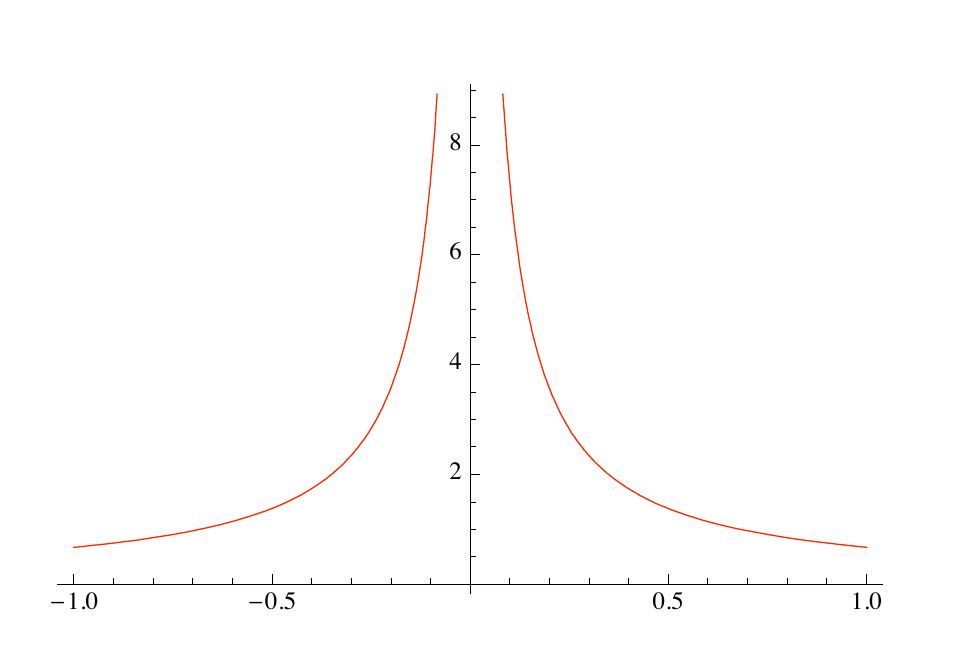}
}
\caption{\footnotesize Representation of the function $g_q(t)$ for $q=(3-\sqrt{5})/2$.}
\label{g_functions}
\end{figure}

Now, from \cite{GazBaG09} and \cite{berg1} a positive measure solution to the moment problem for the sequence 
$\left( q^{-\frac{n(n+1)}{2}}\right)_{n\in \N}$, $0<q<1$,  is given by
\begin{equation}
\label{momqnn0}
q^{-\frac{n(n+1)}{2}} = \int_0^{\infty} t^n g_q(t)\, dt\, ,
\end{equation}
with
\begin{equation*}
\label{momqnn}
g_q(t) = \frac{1}{\sqrt{2 \pi\, \vert \ln{q}\vert}}\, \exp\left\lbrack - 
\frac{\left\lbrack\ln \left(\frac{t}{\sqrt{q}}\right)\right\rbrack^2}{2 \vert \ln q\vert}\right\rbrack\, .
\end{equation*}

Finally, applying the composition formula (\ref{convmel})  
 to the product of  (\ref{momeas}) and (\ref{momqnn0}) yields the solution to the moment problem 
 for the sequence $(x_n!)$:
\begin{equation*}
\label{momsolxn}
x_n! = q^{-\frac{n(n+1)}{2}} \, q^{\frac{n(n+1)}{2}}\, x_n! = \int_0^{\infty} t^n w_q(t)\, dt\, 
\end{equation*}
with the positive measure density given by
\begin{equation*}   
w_q(t) =  \int_0^{\infty}  g_q(t/u)\, \varpi_q(u)\, \frac{du}{u}= (q^{-1}-q)\sum_{j=0}^\infty  \, g_q\left(t\, \frac{q^{-1}-q}{q^{2j}}\right) \mathfrak{E}_q\left(-\frac{q^{2j}}{q^{-1}-q}\right) \, .
\end{equation*}


\section{CS quantization of the complex plane with integer symmetric deformations of the integer numbers}\label{CSqsymmetricdeformations}

As was announced above, we  fix our attention to Case 2 in Section~\ref{Integersasdeformationsofintegers} by considering the sequence of integers 
$\left( x_n=  {}^{[s]}[n ]_{q}\right)$ defined for $n \geq 0$ by 
\eqref{qsdef}
where $q$ ($0<q<1$) is solution of eq. \eqref{qpequatunit+}, i.e, $X^2 -sX +1 = 0$ with integer $s \geq 3$.
From (\ref{unequat}) these integers $ x_n$ obey the recurrence relation 
\begin{equation*}
\label{xnrec1}
x_{n+1} -s x_n + x_{n-1} = 0\, , \quad  x_0 = 0\, , \quad x_1 = 1\,  .
\end{equation*}

In  section~\ref{momentumproblem} we have solved the moment problem for the sequence $(x_n !)_{n\in \N}$, so we can construct a family of CS associated to $(x_n )_{n\in \N}$ according to \eqref{xncs} and solving the identity according to (\ref{rengenquant}). We now proceed with numerical explorations by
choosing the lowest cases $s=3, 4$ and $5$  which correspond to $q= (3 - \sqrt 5)/2,\;  q= 2-\sqrt{3}$ and  $q= (5-\sqrt{21})/2$ respectively. Note that 
$(3 - \sqrt 5)/2 = 1/\tau^2$, where $\tau = (1 + \sqrt 5)/2$ is the golden mean. 
These $q$-numbers ($q= (3 - \sqrt 5)/2$) are Fibonacci numbers  which occupy the odd place in the Fibonacci series (see Table~1). The limit of ${}^s [n]_q$ \eqref{qsdef} when  $q\to 1$ is  $n$.

The explicit expression of the coherent states \eqref{xncs}  associated to these $q$-integers ${}^{[s]}[n ]_{q}$   is
\begin{equation*} 
\label{qxncs}
|v_z\rg _q = \sum_{n=0}^{\infty} \frac{1}{\sqrt{\mathcal{N}_q(\vert z \vert^2)}}\frac{z^n}{\sqrt{{}^{[s]}[n ]_{q}!}}\,  |e_n\rg\, .
\end{equation*}

The CS $|v_z\rg _q $ in the limit $q\to 1$ goes to the standard CS  $|v_z\rg $ \cite{glauber63,klauder63,sudarshan63}
\begin{equation} 
\label{standardcs}
|v_z\rg  = \sum_{n=0}^{\infty} \frac{1}{\sqrt{\exp(\vert z \vert^2)}}\frac{z^n}{\sqrt{n !}}\,  |e_n\rg\, 
\end{equation}
(for which the kets $|e_n\rg$ are given their usual Fock interpretation of number states, i.e. $|e_n\rg \equiv |n\rg $), taking into account that when  $q$ goes to $1$ ${}^{[s]}[n ]_{q}\to n$ and 
$\mathcal{N}_q(\vert z \vert^2)\to \exp{(\vert z \vert^2)}$.

\begin{figure}[t]
\centering
 \subfigure[$d_q(n)$ \hskip6.75cm (b) $\mathcal{N}_q(t)$] {
 \includegraphics[width=0.485\textwidth]{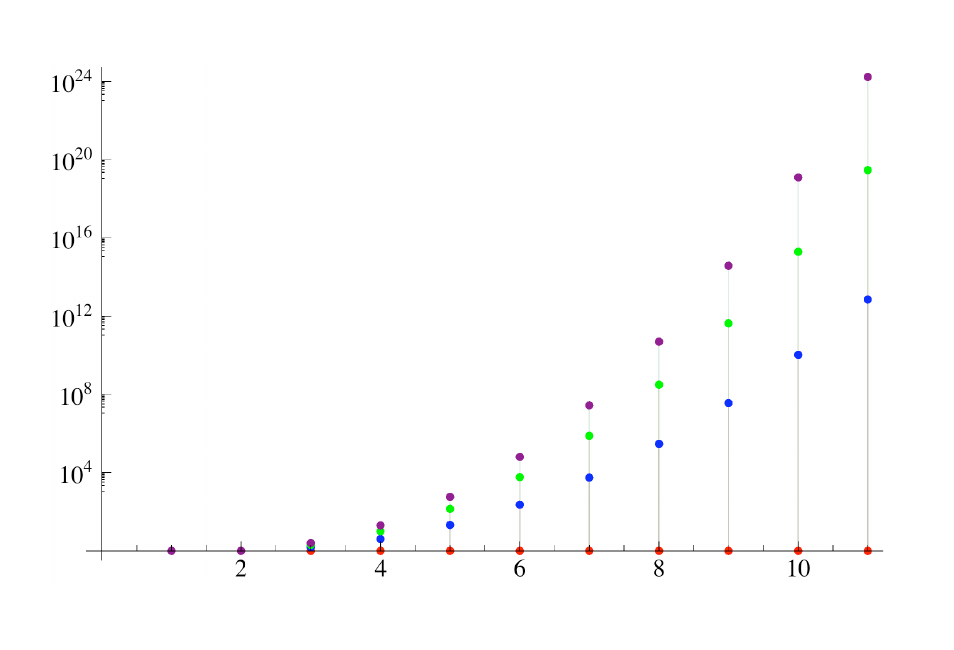}\qquad  
\includegraphics[width=0.485\textwidth]{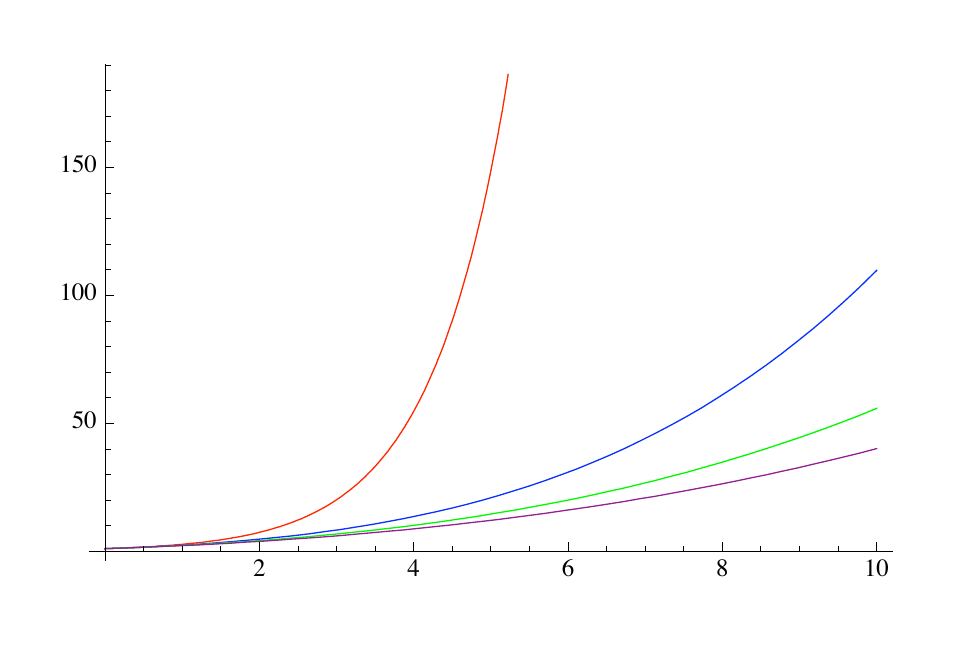}
}
\caption{\footnotesize Representation of  $d_q(n))$ and  $\mathcal{N}_q(t)$ with $t\geq 0$ for  $q=1$ (red),  $(3-\sqrt{5})/2$ (blue), $2-\sqrt{3}$ (green), $(5 - \sqrt{21})/2$ (purple).}
\label{ratio_functions}
\end{figure}

In the following we  study the physical properties of these $q$-CS for different values of $q$ and we compare with the
standard CS. A graphical way to see this limit is to consider  the ratio-function
\[
d_q(n)=\frac{{}^{[s]}[n ]_{q}!}{n!}
\]
and display it versus $n$ for different values of $q$ \cite{quesne02} (see Fig.~\ref{ratio_functions}a). 
Also in figure~\ref{ratio_functions}b we represent the normalization function $\mathcal{N}_q(t)$ \eqref{seqexp1} of the CS versus $t$ for positive values of $t$ and different values of $q$.

\begin{figure}[t]
\centering
 \subfigure [$\vert z\vert =4$ (red), $5$ (blue), $6$ (green) and $q=(3-\sqrt{5})/2$. \hskip.335cm (b) $\vert z\vert =4,\;q=1$ (yellow), 
 $(3-\sqrt{5})/2$ (red),  \newline . \; \hfill
 $2-\sqrt{3}$ (blue),
 $(5 - \sqrt{21})/2 $ (green).]{
 \includegraphics[width=0.485\textwidth]{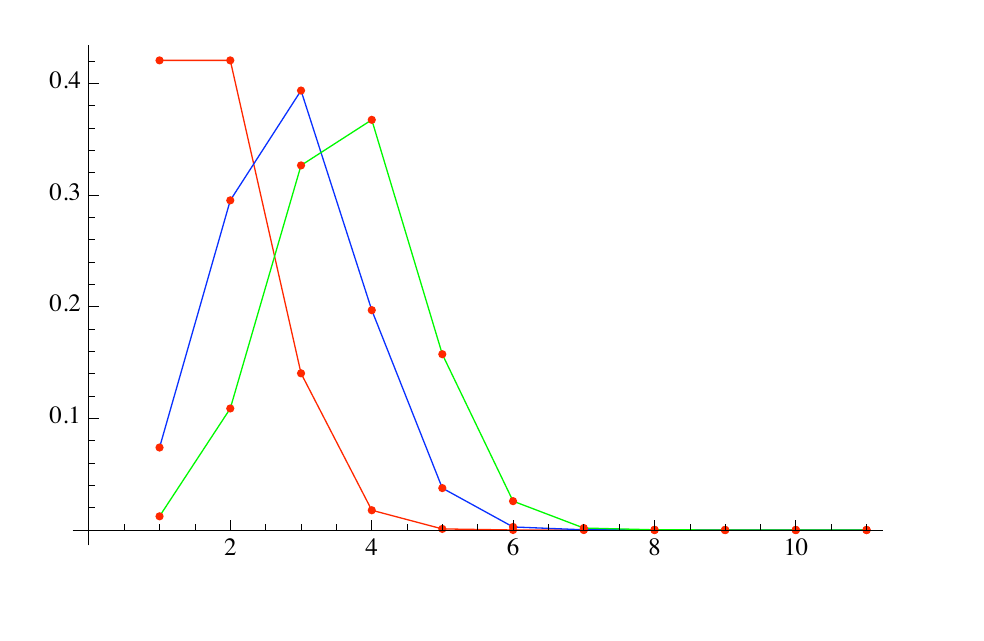}\qquad  
\includegraphics[width=0.485\textwidth]{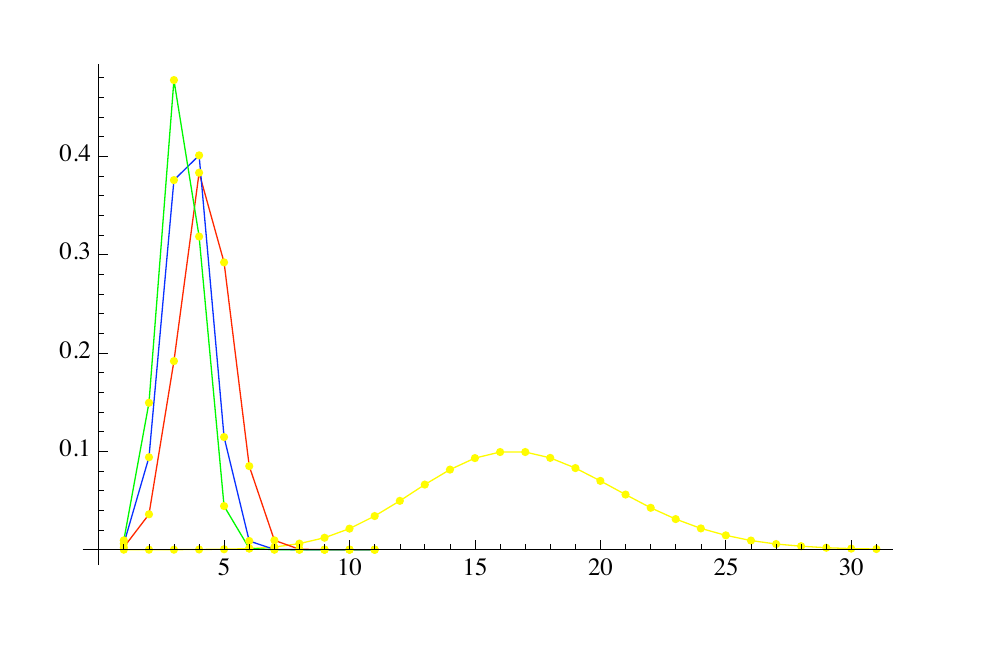}
}
\caption{\footnotesize Representations of Poisson-like distributions $\rho_q(n,\vert z\vert)$ versus $n$ for different values of the parameters $\vert z\vert $ and $q$.}
\label{poisson_distributions}
\end{figure}

\begin{figure}[h]
\centering
\includegraphics[scale=.85]{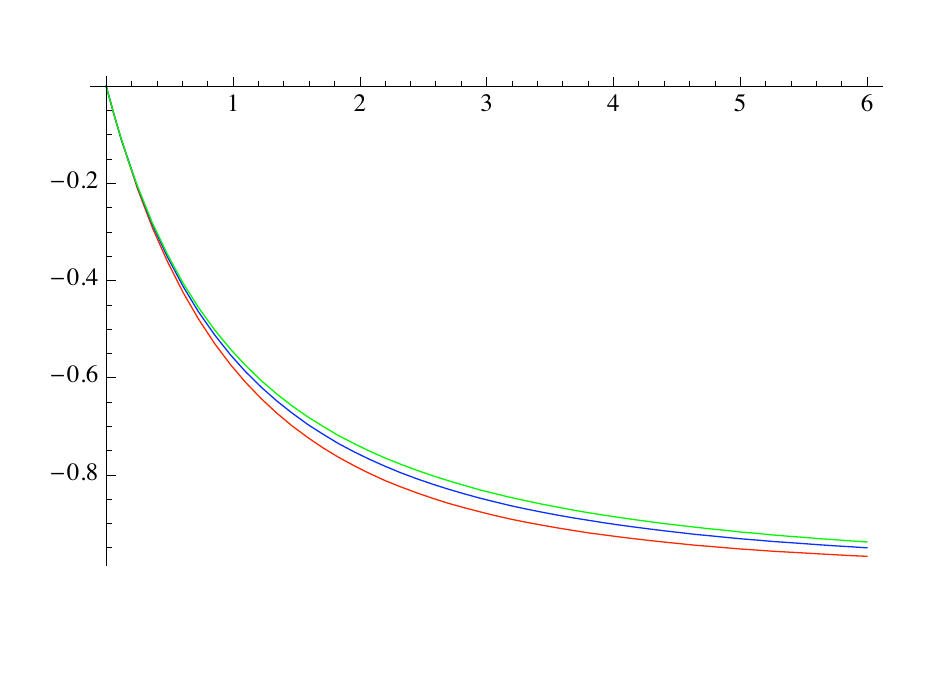}
\vskip-0.75cm

\caption{\footnotesize  The Mandel parameter $Q_q(\vert z \vert^2)$ versus $\vert z \vert$  for $ q=(3-\sqrt{5})/2$ (red), 
$2-\sqrt{3}$ (blue), $(5 - \sqrt{21})/2$ (green).} \label{fig_mandel}
\end{figure}

Let us suppose that the state $|e_n\rg $ corresponds to the state of $n$ bosons $ |n\rg $. Then the probability of finding $n$ bosons in the $q$-CS $|v_z\rg _q  $  is given by the 
$q$-Poisson distribution 
\begin{equation*}
\rho_q(n,\vert z\vert)=\dfrac{\vert z \vert^{2n}}{\mathcal{N}(\vert z \vert^2)\;x_n!} ,
\end{equation*}
whose limit when $q\to 1$ is the standard Poisson distribution appearing in the standard CS. 
In figures~\ref{poisson_distributions} we display the $q$-Poisson distribution for different values of $\vert z\vert $ and $q$. In particular, in figure~\ref{poisson_distributions}b we also display the standard Poisson distribution (in yellow) and  notice that the $q$-Poisson distributions are left displaced with respect to the standard Poisson distribution. It shows that 
$\rho_q(n,\vert z\vert)$ is a sub-Poissonian distribution. The Mandel parameter \cite{mandel}
\begin{equation*}
Q_q(\vert z \vert^2)=\frac{(\Delta X_N)^2-\lg X_N\rg}{\lg X_N\rg}=\frac{\lg X_N^2\rg-\lg X_N\rg ^2-\lg X_N\rg}{\lg X_N\rg}
\end{equation*}
measures the deviation of a Poisson-like distribution from the Poisson distribution. The Mandel parameter for the Poisson distribution vanishes since for standard CS $(\Delta N)^2= \lg N\rg$, where $N$ is the number operator; if  
$Q_q(\vert z \vert^2)<0$ we have a  sub-Poissonian distribution, i.e. there is an antibunching effect;  finally if  
$Q_q(\vert z \vert^2)>0$ the $q$-Poisson distribution is a super-Poissonian distribution, i.e. the bunching effect appears.
For our $q$-CS the Mandel parameter is negative 
(see Fig.~\ref{fig_mandel}).

\begin{figure}[h]
\centering
\includegraphics[scale=.85]{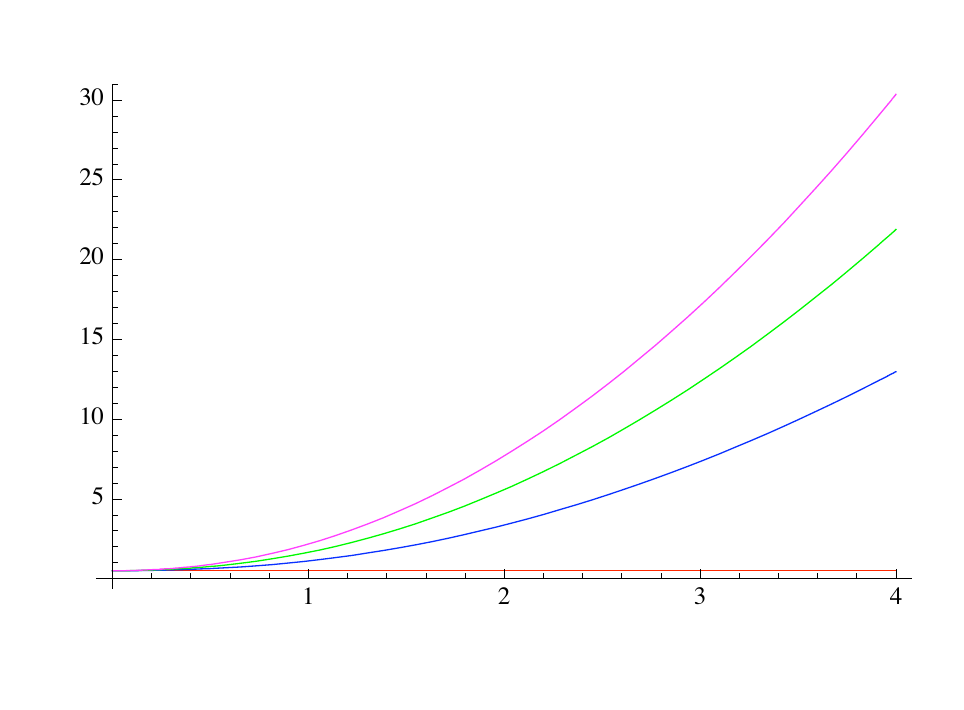}
\vskip-0.75cm

\caption{\small $(\Delta_{v_z} Q)^2$ versus $|z|$ for $ q=1$ (red),  $(3-\sqrt{5})/2$ (blue),  $2-\sqrt{3}$ (green),  $(5 - \sqrt{21})/2$ (purple).} \label{fig_variance}
\end{figure}

Expression \eqref{uncert} shows us that these $q$-CS 
are intelligent states for the operators $Q$ and $P$ \eqref{QP-operators}. The variances of $Q$ and $P$ read respectively:
\[
(\Delta_{v_z} Q)^2=(\Delta_{v_z} P)^2= \frac 12 \vert  \lg v_z |[Q,P]| v_z\rg \vert =  \frac 12\left\vert \frac{1}{\mathcal{N}(\vert z\vert^2)} + (s-1)\vert z\vert^2 - \vert z\vert^4\left\lg \frac{1}{x_{n+2}}\right\rg\right\vert \, .
\]
This quantity is $\geq 1/2$ for all $z\in \C$ since $x_{N+1}-x_N (= (s-1) x_{N}- x_{N-1})\geq  I_d$.
Only for $z=0$, i.e. for the vacuum state $| v_0\rg=| e_0\rg$, we have:
\[
(\Delta_{v_0} Q)^2=(\Delta_{v_0} P)^2= \frac 12 \vert  \lg v_0 |[Q,P]| v_0\rg \vert =1/2\, .
\]
So, these operators have no minimum-uncertainty state in the sense of ordinary quantum mechanics, at the exception of the vacuum,  in such a way that the vacuum uncertainty product provides a global lower bound (see Fig.~\ref{fig_variance}).
According to  these results squeezing does not  occur in $|v_z\rg _q$ for any $q$ and $z$.

The $q$-deformed boson creation and annihilation operators ($a,\, a^\dagger$) \eqref{stoper1} and \eqref{stoper2} can be related to the standard  boson creation and annihilation operators ($b,\; b^\dagger$) by
\[
a^\dagger=\sqrt{\frac{x_{n+1}}{n+1}}\; b^\dagger ,\qquad
a=\sqrt{\frac{x_n}{n}}\; b .
\]
Thus, standard $n$-boson states $|e_n\rg$ can be rewritten as $n$-$q$-deformed-boson states as
 \[
|e_n\rg 
 =\frac{1}{\sqrt{n!}}(b^\dagger)^n |e_0\rg  
=\frac{1}{\sqrt{x_n!}}(a^\dagger)^n |e_0\rg .
\] 
Hence,   the $q$-CS $|v_z\rg$ can also expressed in terms of   $n$-bosons operators  by
\[
|v_z\rg = \sum_{n=0}^{\infty} \frac{1}{\sqrt{\mathcal{N}(\vert z \vert^2)}}\frac{z^n}{{x_n!}}\,(a^\dagger)^n  |e_0\rg\,
=\sum_{n=0}^{\infty} \frac{1}{\sqrt{\mathcal{N}(\vert z \vert^2)}}\frac{z^n}{\sqrt{x_n! \;n!}}\,(b^\dagger)^n  |e_0\rg\, .
\]

\begin{figure}[h]
\centering
\includegraphics[scale=.65]{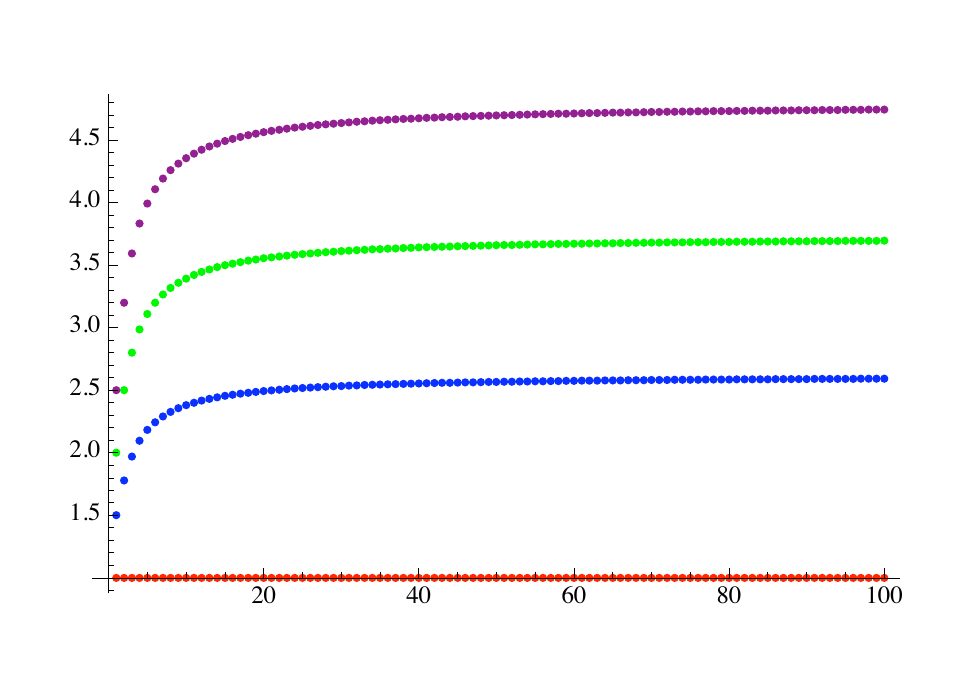}
\vskip-0.75cm

\caption{\small Characteristic functional  for $ q=1$ (red),  $(3-\sqrt{5})/2$ (blue),  $2-\sqrt{3}$ (green),   $(5 - \sqrt{21})/2$ (purple).} \label{fig_characteristic}
\end{figure}

We can compare the  properties of the $q$-CS with those of the standard CS by using the characteristic functional \cite{solomon94}  (see Fig.~\ref{fig_characteristic})
\[
\rho_q(n)=\frac{^{[s]}[n +1]_{q}/ ^{[s]}[n ]_{q} }{(n+1)/n}
\] 
Obviously, in our case,  $\rho_q(n)>1$ for any $q$ (see Fig.~\ref{fig_characteristic}).
 
\begin{figure}[h]
\centering
\includegraphics[scale=.65]{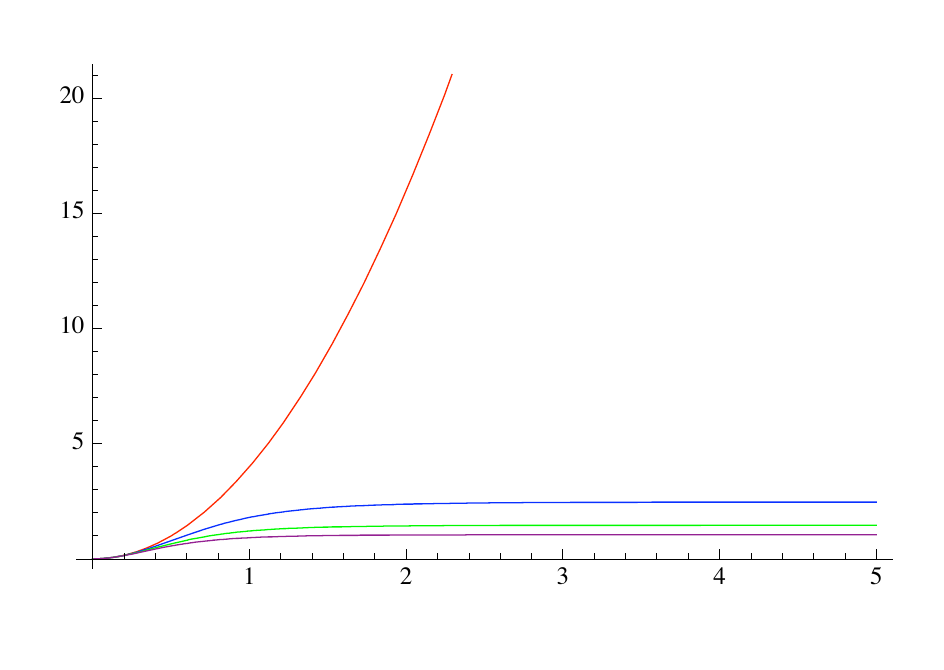}
\vskip-0.75cm

\caption{\small Signal-to-quantum noise ratio  versus $|z|$ for $ q=1$ (red), $(3-\sqrt{5})/2$ (blue), $2-\sqrt{3}$ (green), $(5 - \sqrt{21})/2$ (purple).} \label{fig_nose}
\end{figure}

Signal-to-quantum noise ratio  for these $q$-deformed photons is determined by the relation
\[
\sigma_q=\frac{\lg Q \rg _q^2}{(\Delta Q)_q^2} .
\]		
Note that $\lg Q \rg _q=\sqrt{2} \Re z $.
 Fig.~\ref{fig_nose} shows that the signal-to-quantum noise ratio  for $q$-deformed photons is not enhanced over the standard  photon case (in red). 

\begin{figure}[t]
\centering
 \subfigure [Standard coherent states. \hskip2.95cm (b) $q$-Pisot coherent states ($q=(3-\sqrt{5})/2$).]{
 \includegraphics[width=0.485\textwidth]{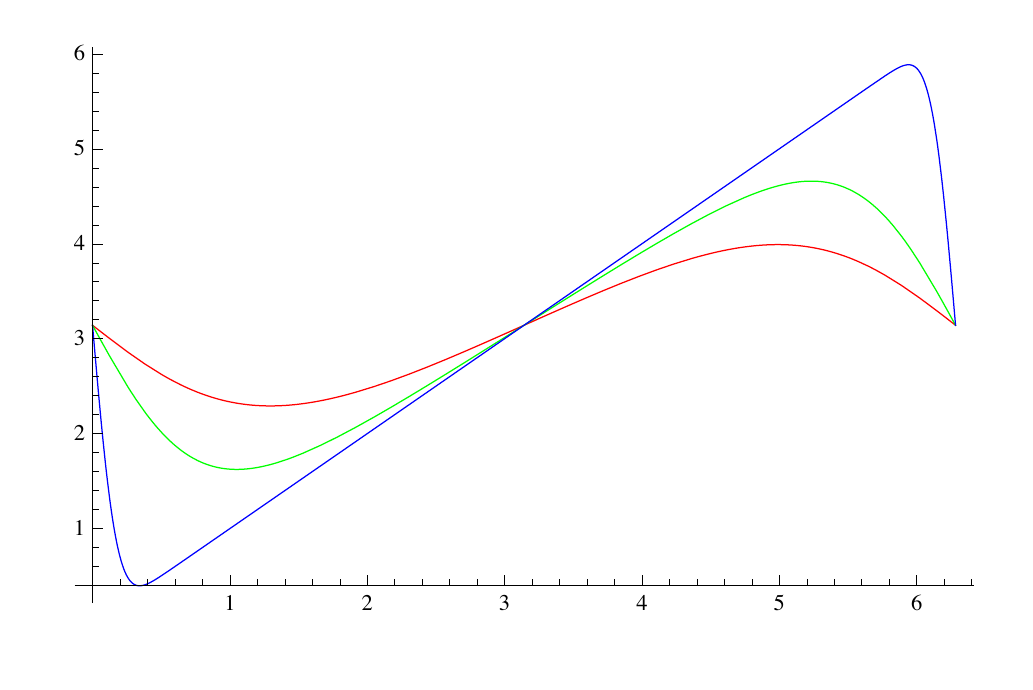}\qquad  
\includegraphics[width=0.485\textwidth]{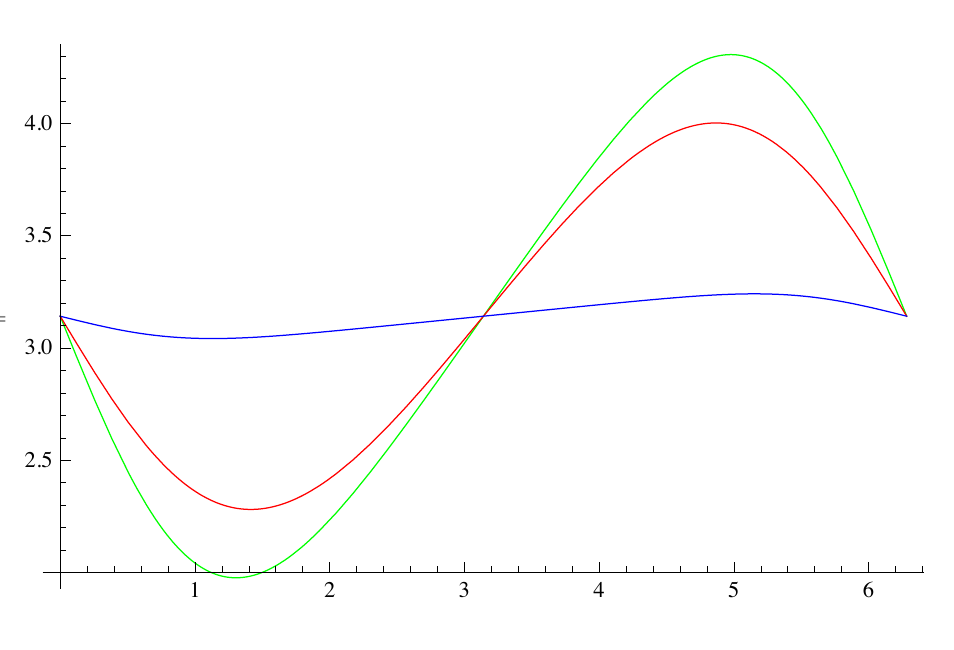}
}
\caption{\footnotesize Lower symbol of the angle operator for different values of the parameters $|z|=0.5$ (red), $1$ (green), $5$ (blue) and $\theta \in [0,2 \pi)$.}
\label{low_symbol-angle1}
\end{figure}

\begin{figure}[t]
\centering
 \subfigure [$|z|=0.5$. \hskip6.5cm (b) $|z|=2$.]{
 \includegraphics[width=0.485\textwidth]{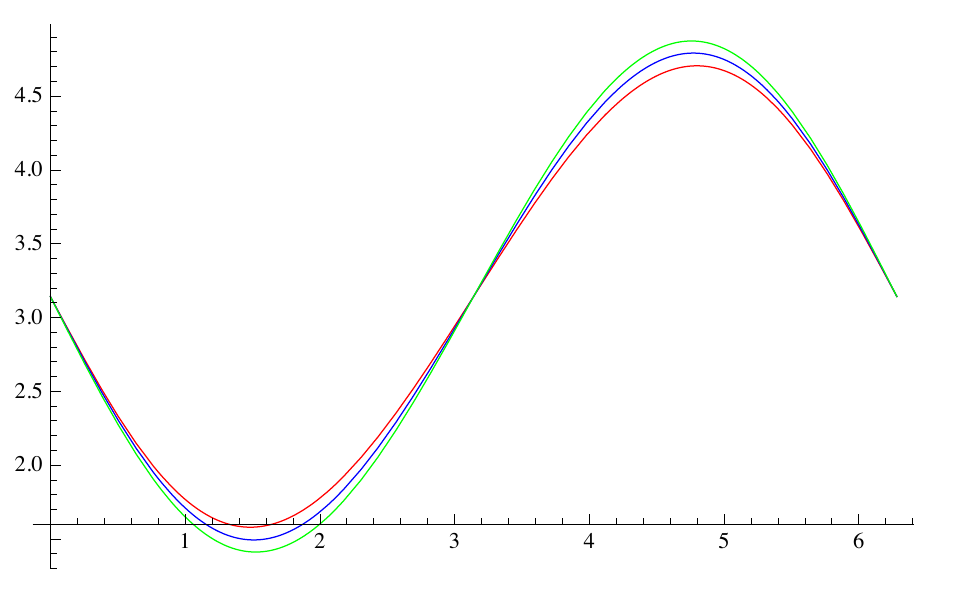}\qquad  
\includegraphics[width=0.485\textwidth]{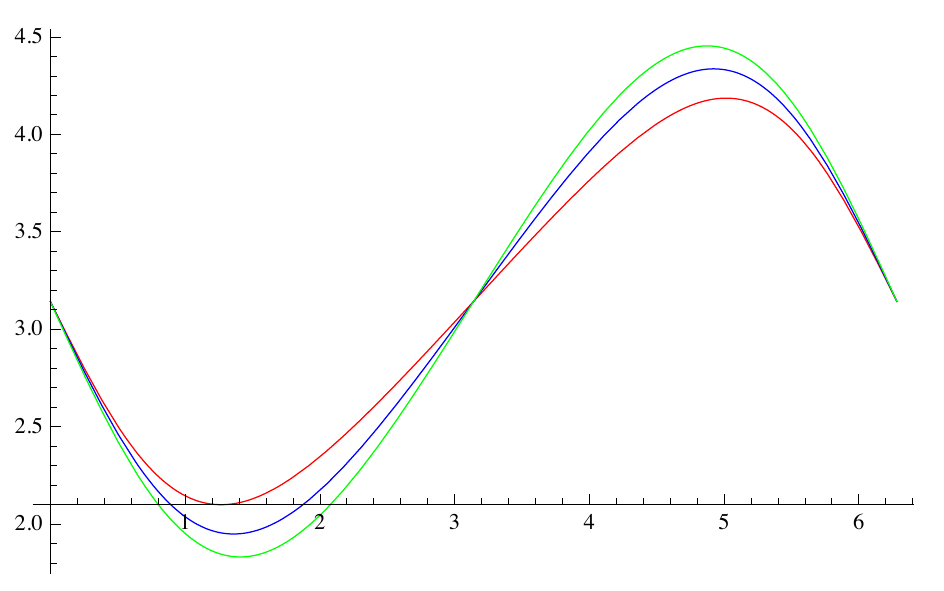}
}
\caption{\footnotesize Lower symbol of the angle operator for   $ q=(3-\sqrt{5})/2$ (red),  $2-\sqrt{3}$ (blue), $(5 - \sqrt{21})/2$ (green) and 
$\theta \in [0,2 \pi)$..}
\label{low_symbol-angle2}
\end{figure}

\begin{figure}[t]
\centering
 \subfigure [Standard coherent states. \hskip1.5cm (b) $q$-Pisot coherent states ($q=(3-\sqrt{5})/2$).]{
 \includegraphics[width=0.485\textwidth]{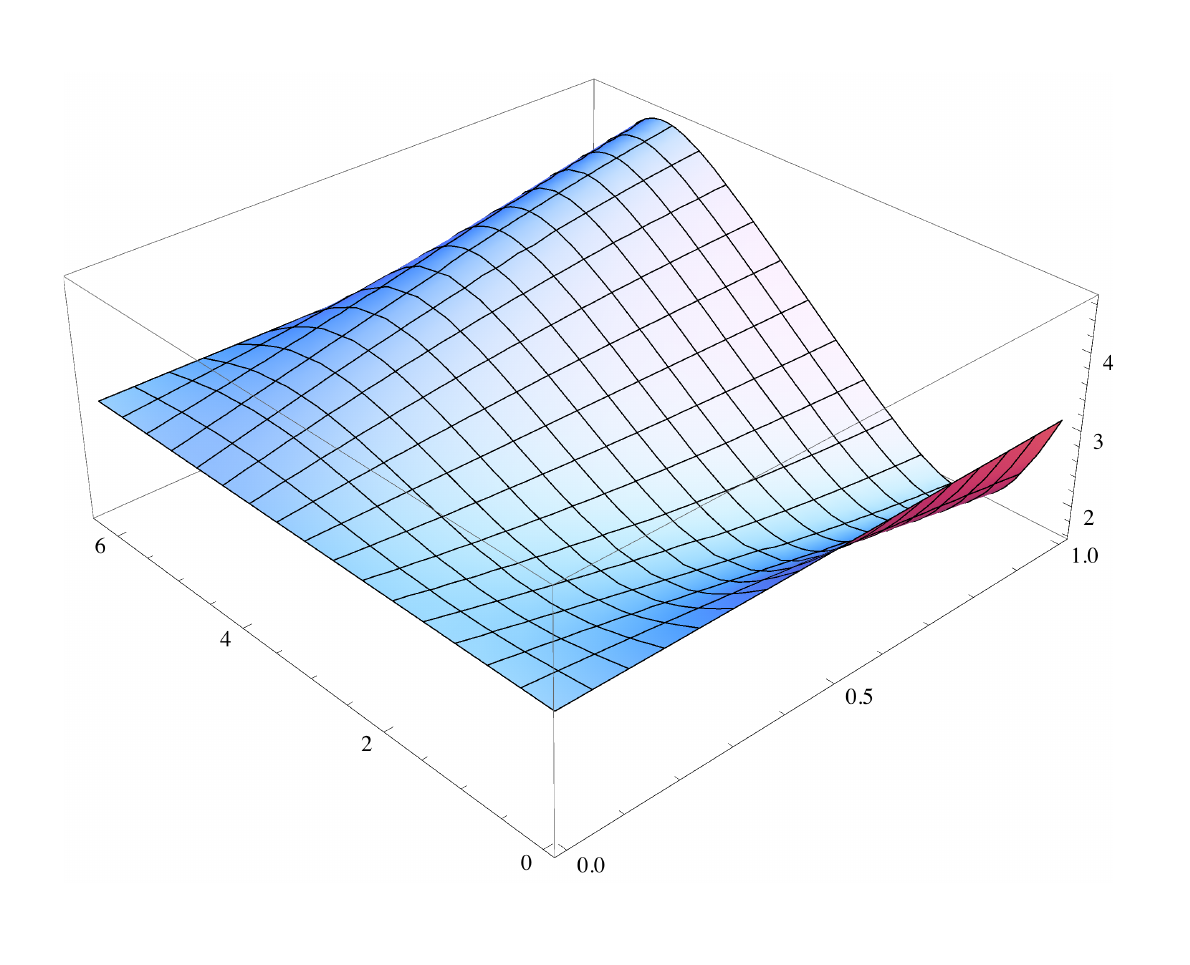}\qquad  
\includegraphics[width=0.485\textwidth]{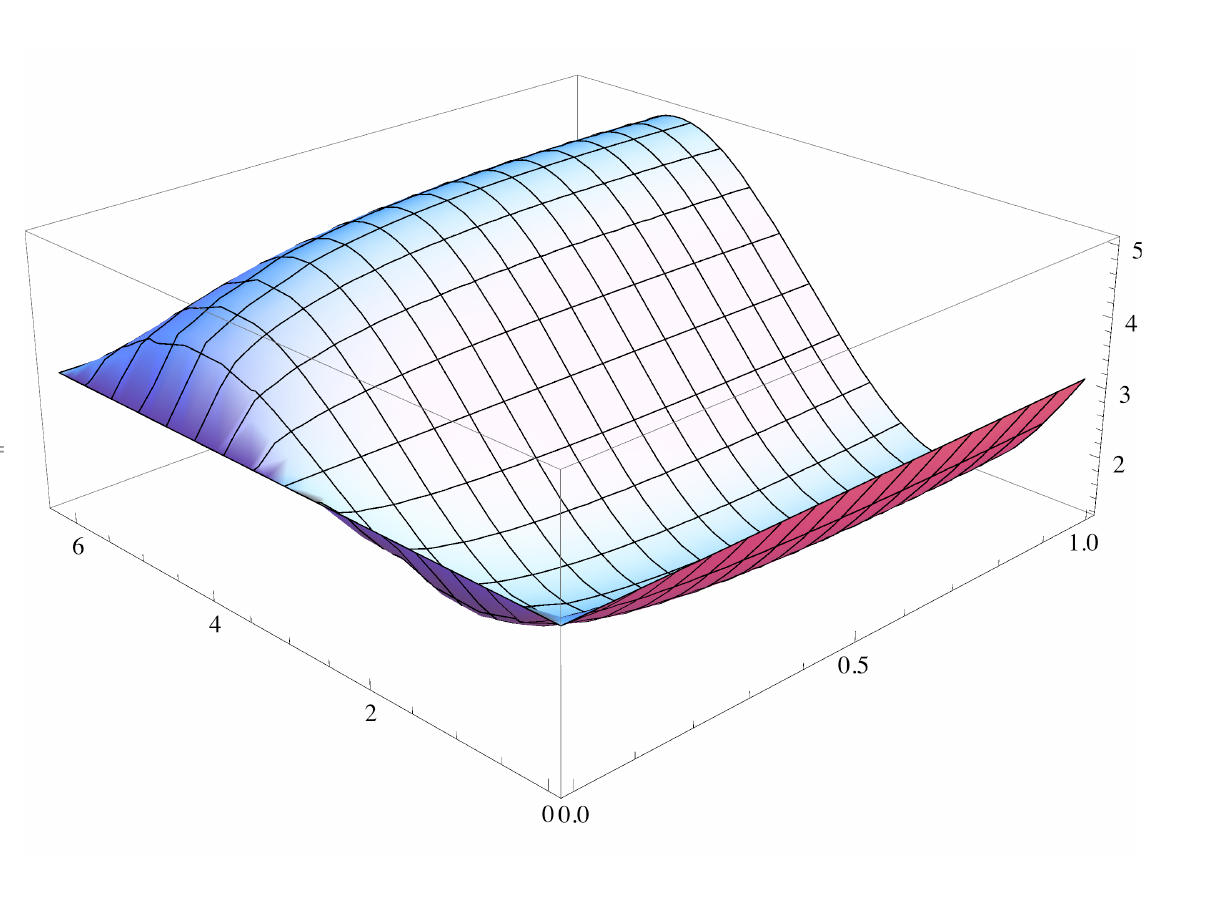}
}
\caption{\footnotesize Lower symbol of the angle operator for   $|z|\in[0,1]$ and $\theta \in [0,2 \pi)$.}
\label{low_symbol-angle3}
\end{figure}

\begin{figure}[t]
\centering
 \subfigure [$k=1$ (blue), $k=2$ (purple), $k=4$ (green),  \hskip1.5cm (b): $k=1$;   $q=(3-\sqrt{5})/2$ (blue),  $q=2-\sqrt{3}$ 
  $k=10$ (red).\hskip7.75cm (red).]{
 \includegraphics[width=0.485\textwidth]{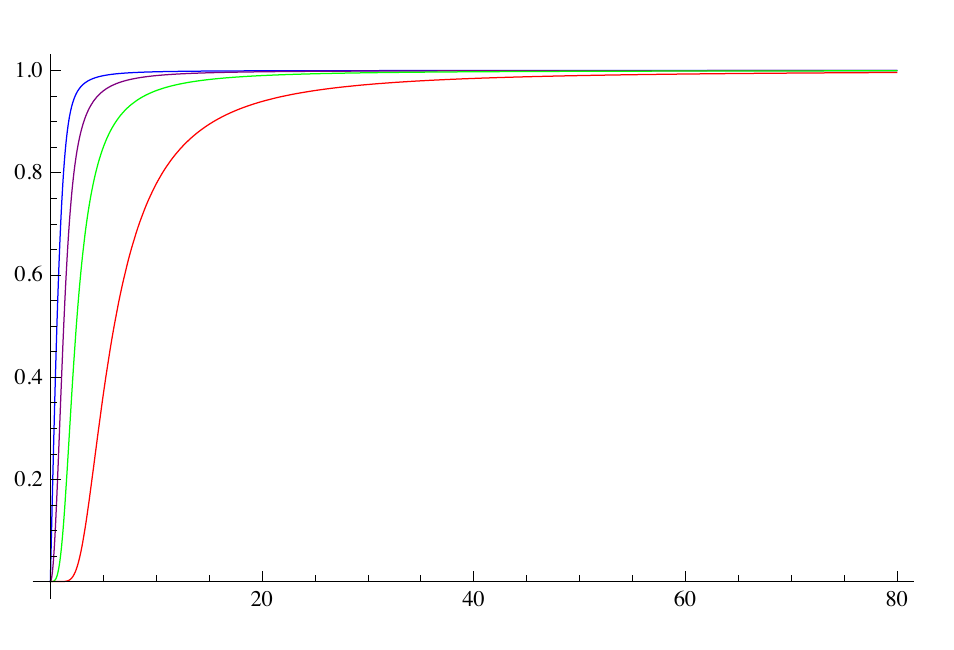}\qquad  
\includegraphics[width=0.485\textwidth]{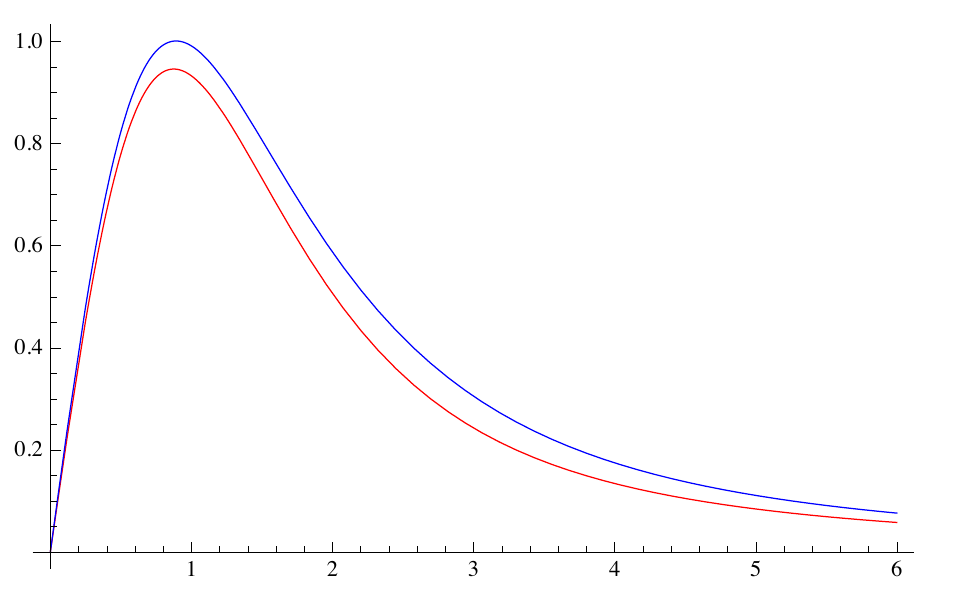}
}
\caption{\footnotesize Representation of the function $d_k(r)$ \eqref{dkr}:   (a) for $q=1$ and 
(b) for $q\neq 1$.}
\label{fig_dkr}
\end{figure}

\begin{figure}[t]
\centering
 \subfigure [$k=1$ (red), $k=2$ (blue), $k=4$ (green)   \hskip1.5cm (b): $k=2$;   $q=(3-\sqrt{5})/2$ (red),  $q=2-\sqrt{3}$ (blue) 
  and $q=(3-\sqrt{5})/2$. \hskip5.7cm and $q=(5-\sqrt{21})/2$  (red).]{
 \includegraphics[width=0.485\textwidth]{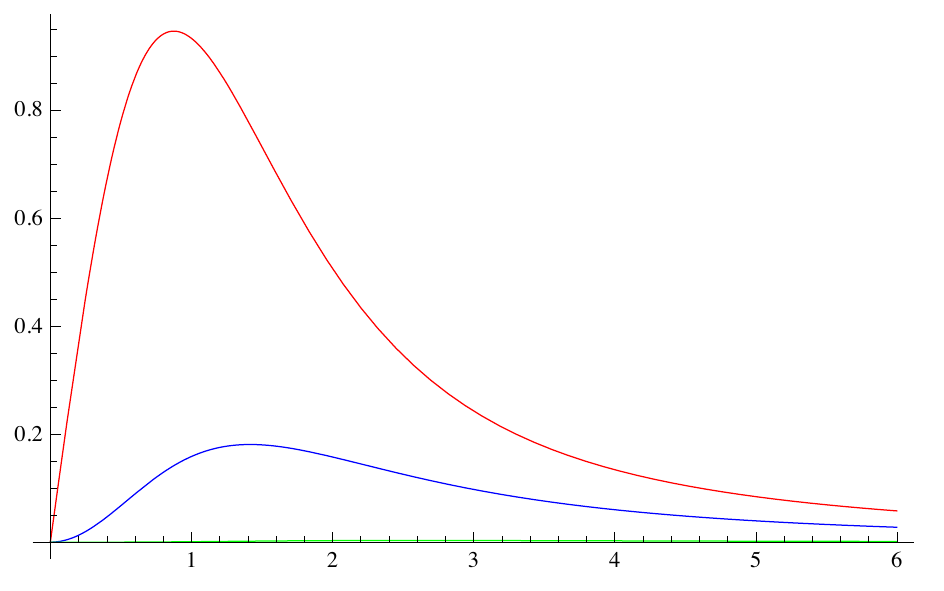}\qquad  
\includegraphics[width=0.485\textwidth]{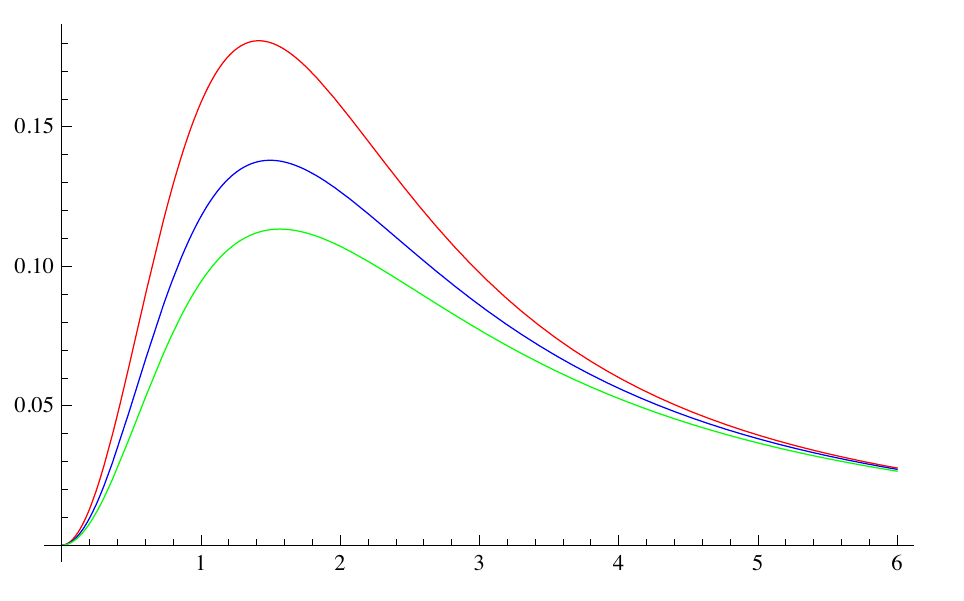}
}
\caption{\footnotesize Representation of the function $d_k(r)$ \eqref{dkr}   for $q\neq 1$.}
\label{fig_dkrbis}
\end{figure}

In relation with the $q$-Pisot  coherent state quantization of angle (angle operator $A_\theta$  \eqref {angleop1}) we  display (Figs.~\ref{low_symbol-angle1}, \ref{low_symbol-angle2} and \ref{low_symbol-angle3}) the  lower symbol of  $A_\theta$ 
($ \check{\theta}(z,\bar z)$)  whose explicit expression is given by (\ref{angleop2}). In order to make comparisons we also display  in some of these figures the angle lower-symbol for the standard coherent states (\ref{standardcs}). We easily check that the function $ \check{\theta}(z,\bar z)$ tends to the classical angle function as $\vert z \vert \to \infty$. Effectively, it is enough to se that the function $d_k(r)$ goes to 1 when $r=|z|=r$ goes to $\infty$ (and $q=1$) (see Fig.~\ref{fig_dkr} and~\ref{fig_dkrbis}).
%

\begin{figure}[t]
\centering
 \subfigure [$q=(3-\sqrt{5})/2$. \hskip5cm (b)  $q=2-\sqrt{3}$.]{
 \includegraphics[width=0.45\textwidth]{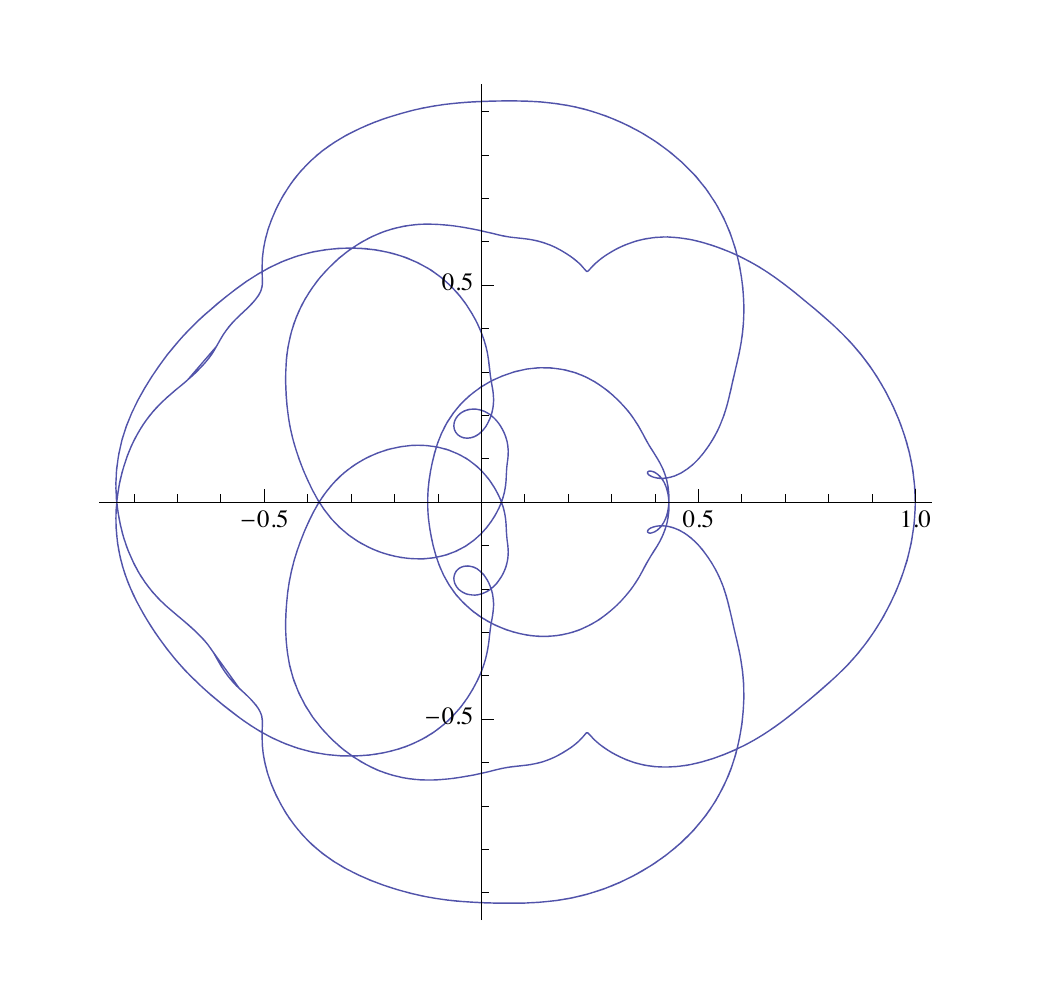}\qquad  
\includegraphics[width=0.45\textwidth]{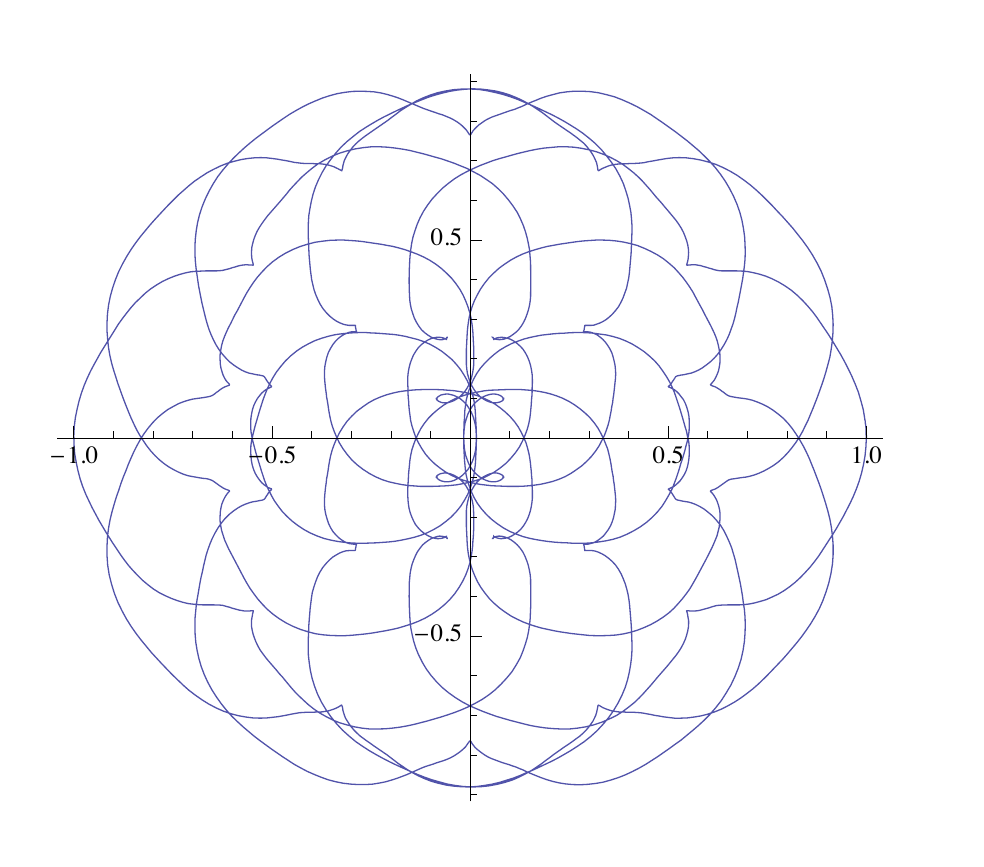}
}
\caption{\footnotesize Plots of $\Im \check{z}\left(  t\right)$ versus $\Re \check{z}\left(  t\right)$  for different values of $q$ and $0\leq t\leq 8 \pi$.}
\label{time_evolution-q23}
\end{figure}
\begin{figure}[t]
\centering
 \subfigure [
$q=(5 - \sqrt{21})/2 $.  \hskip5cm (b)  ]{
\includegraphics[width=0.45\textwidth]{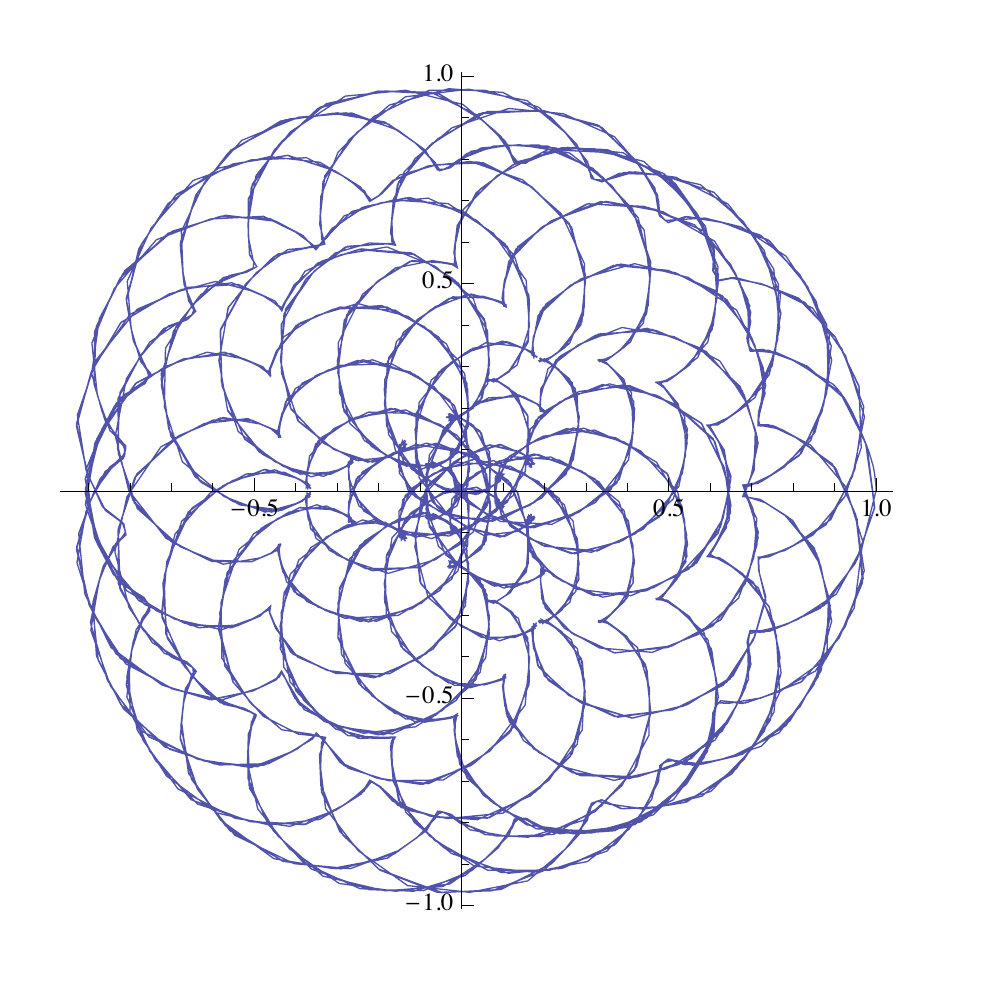}\qquad  
\includegraphics[width=0.45\textwidth]{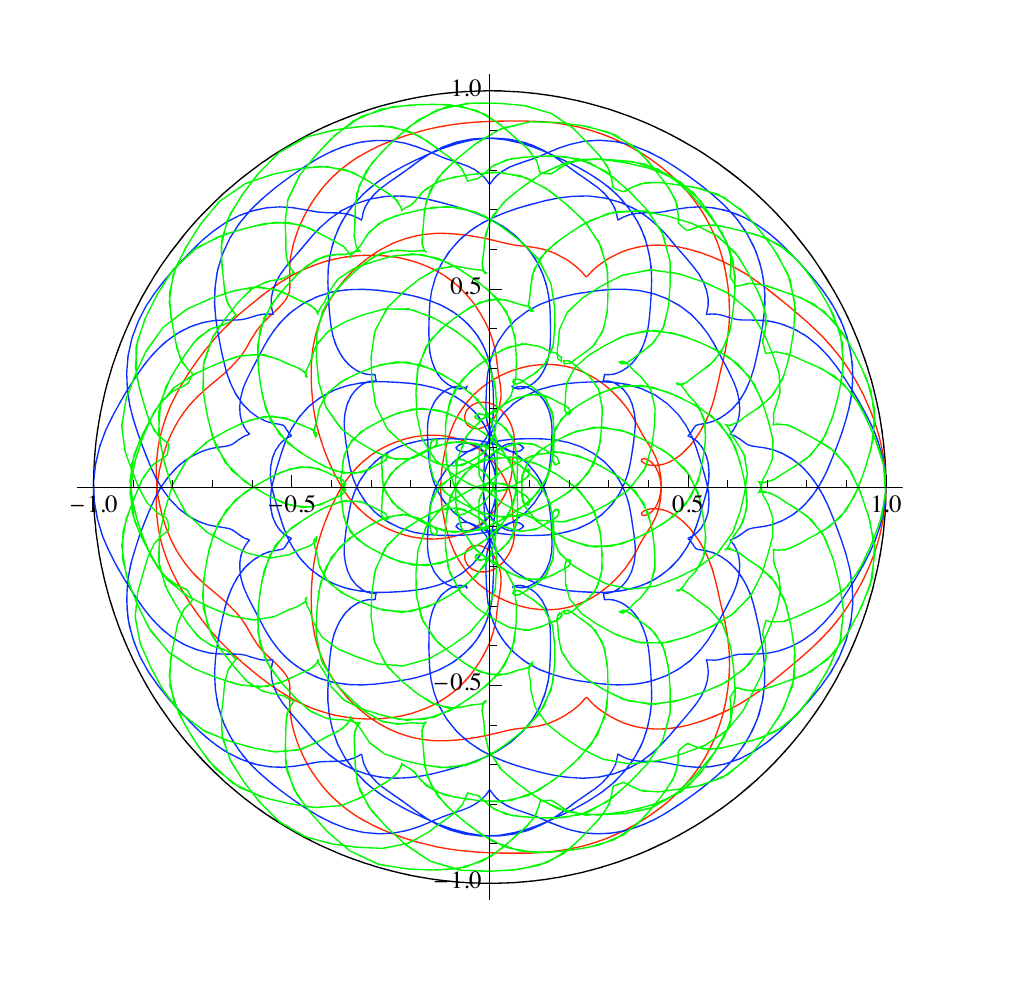}
}
\caption{\footnotesize Plots of $\Im \check{z}\left(  t\right)$ versus $\Re \check{z}\left(  t\right)$  for different values of $q$ and $0\leq t\leq 8 \pi$. In Fig.~\ref{time_evolution-q25}(b) we plot  the previous three figures together with the corresponding to $q=1$ which is a circle of radius 1.}
\label{time_evolution-q25}
\end{figure}

\begin{figure}[t]
\centering
 \subfigure [
Standard coherent states.  \hskip2.5cm (b) $q$-Pisot coherent states ($q=(3-\sqrt{5})/2$).]{
\includegraphics[width=0.45\textwidth]{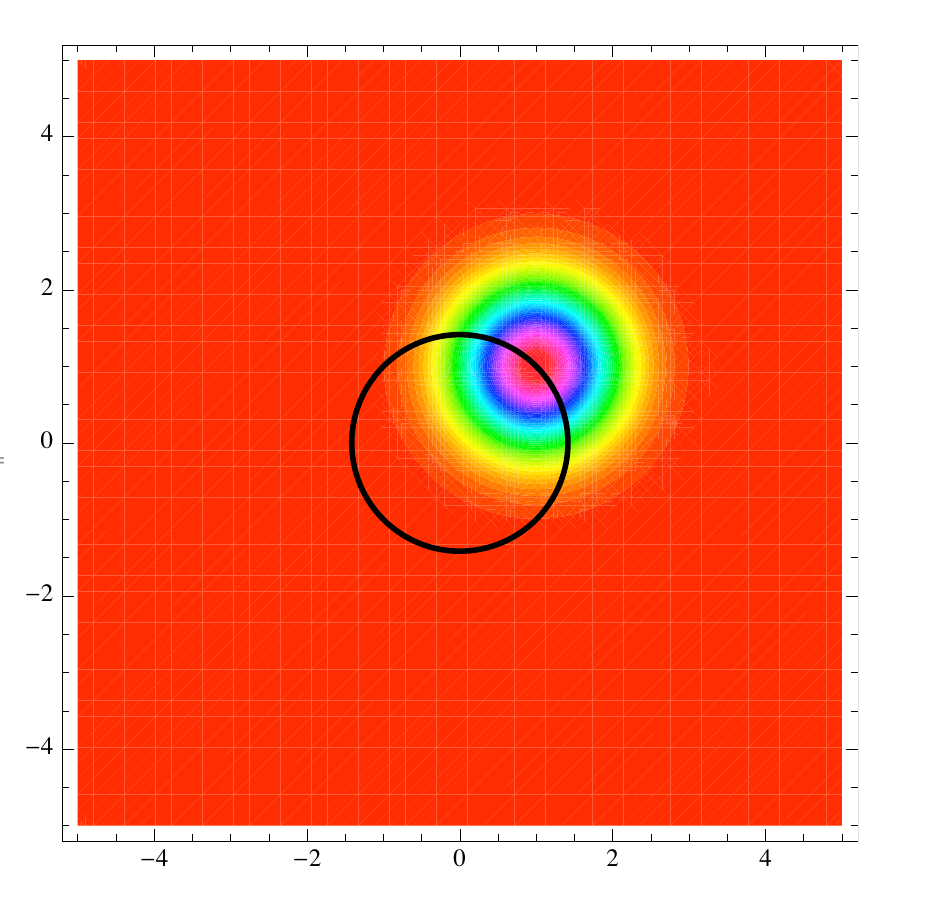}\qquad  
\includegraphics[width=0.45\textwidth]{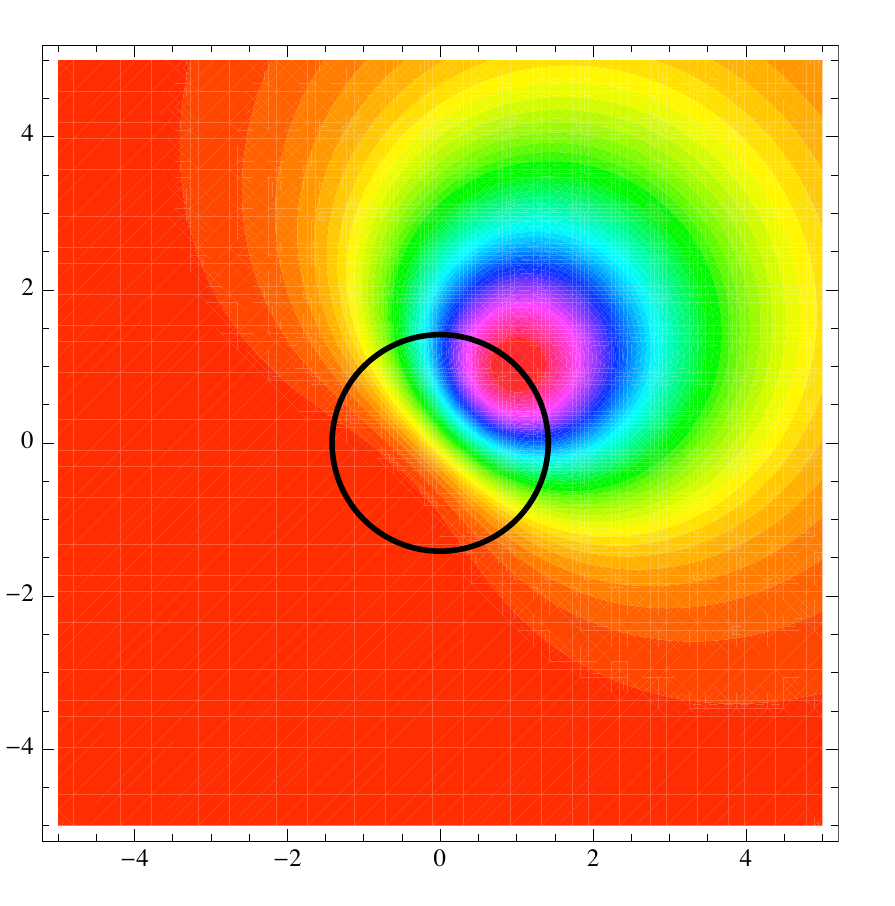}
}
\caption{\footnotesize Phase space distribution $\rho_{v_{z_0}}(z)$ for the particular state $|v_{z_0}\rangle$ with 
$z_0=(1,1)$. The black circle corresponds to the expected phase space given by \eqref{timeev}.}
\label{semiclassical-ro}
\end{figure}

\begin{figure}[t]
\centering
 \subfigure [
$q$-Pisot coherent states ($q=2-\sqrt{3}$)\hskip1.5cm  (b) $q$-Pisot coherent states ($q=(5 - \sqrt{21})/2 )$. ]{
\includegraphics[width=0.45\textwidth]{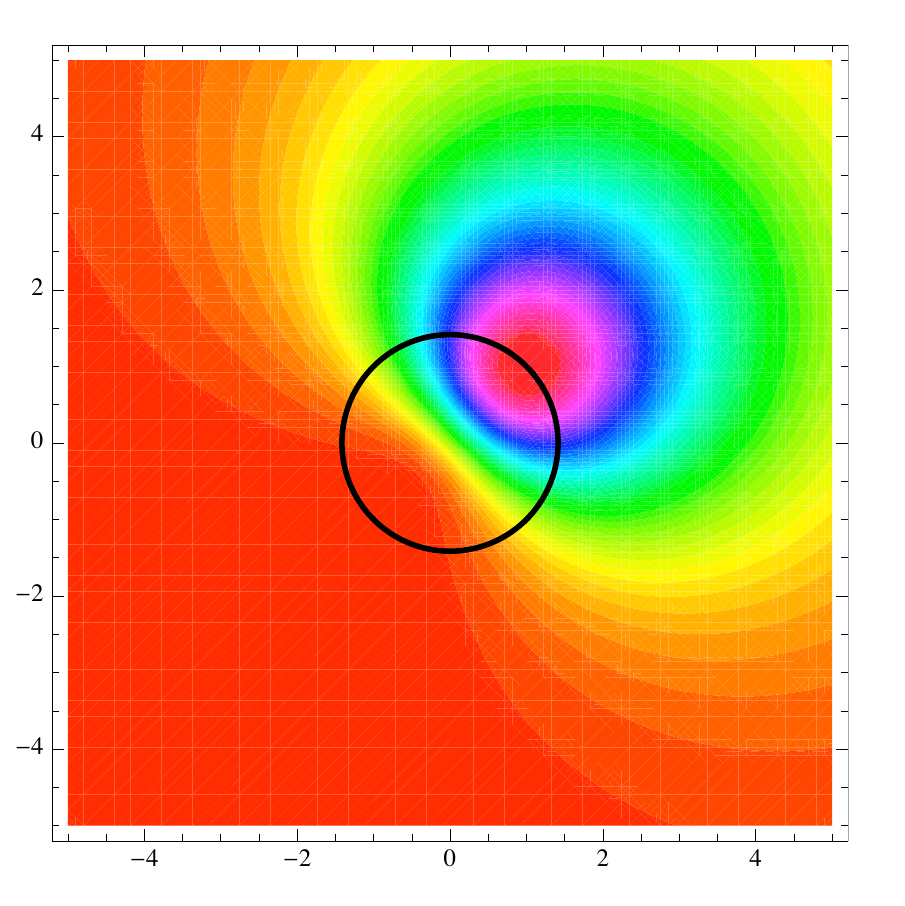}\qquad  
\includegraphics[width=0.45\textwidth]{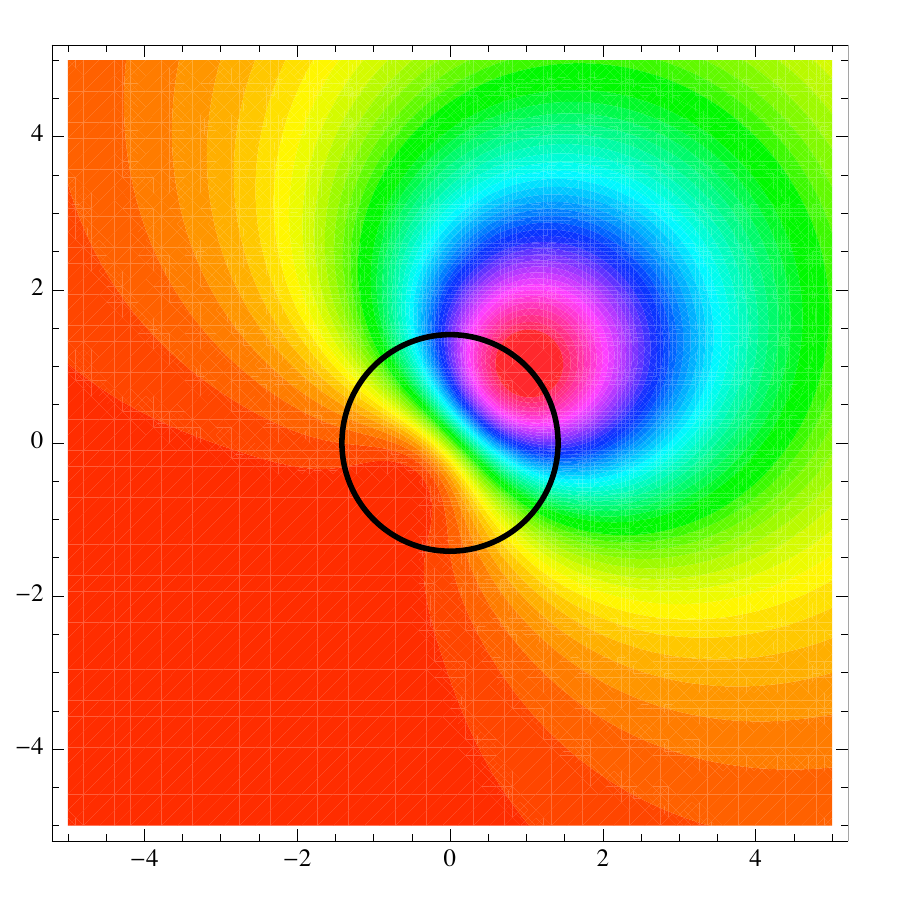}
}
\caption{\footnotesize Phase space distribution $\rho_{v_{z_0}}(z)$ for the particular state $|v_{z_0}\rangle$ with 
$z_0=(1,0)$. The black circle corresponds to the expected phase space given by \eqref{timeev}.
}
\label{q24_semiclassical}
\end{figure}

Another domain of interest is the behavior of the $q$-Pisot coherent states
 in phase space. For this purpose we firstly  study   the trajectories in phase space, i.e., 
 the time evolution of $a = A_z$ (\ref{stoper1}), i.e.  $A_z(t)=e^{-i\hat{H}t}\,  A_z \, e^{i\hat{H}t}$, corresponding to the
classical phase space point $
z= (\q+i\p)/\sqrt{2}$, which is given by the 
mean value in coherent states $|v_z\rangle$, 
$\check{z}\left(  t\right) = \left\langle v_z\right\vert A_z(t)\left\vert
v_z\right\rangle $ \eqref{timeev}.  In Figs.~\ref{time_evolution-q23} and \ref{time_evolution-q25}   we plot  $\Im \check{z}\left(  t\right)$ versus $\Re \check{z}\left(  t\right)$  for different values of $q$ and $0\leq t\leq 8 \pi$. Note that since the $q$ numbers involved in these CS are integers the phase space trajectories are periodic with period $2\pi$.
We show  the phase space distribution $\rho_{v_{z_0}}(z)$ \eqref{probdensphs} in Fig.~\ref{semiclassical-ro} with the fixed state $|v_{(1,1)}\rangle$ and in Fig.~\ref{q24_semiclassical}  for the initial CS state $z_0=(1,0)$ and different values of 
$q$.

\section{Conclusions}

\begin{figure}[t]
\centering
 \subfigure [$q=\sqrt{2}$. \hskip5cm (b)  $q=e $.]{
 \includegraphics[width=0.45\textwidth]{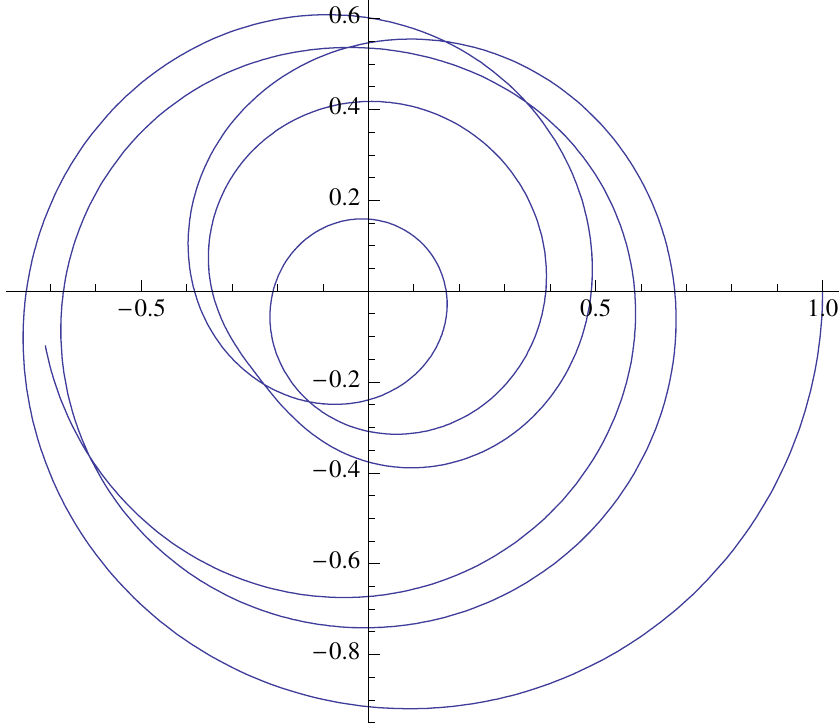}\qquad  
\includegraphics[width=0.45\textwidth]{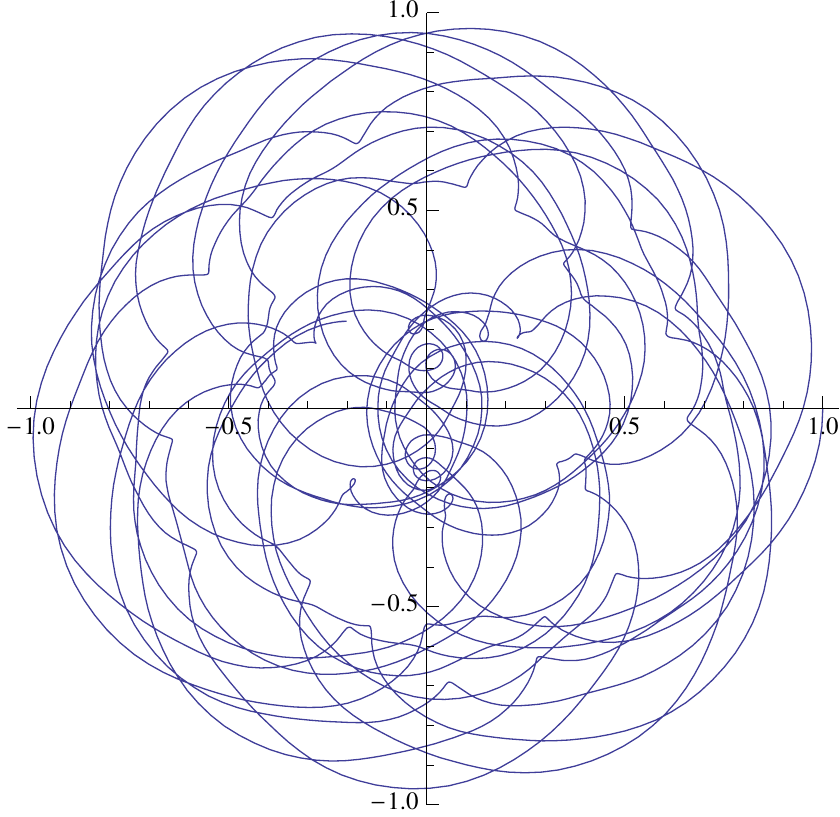}
}
\caption{\footnotesize Plots of $\Im \check{z}\left(  t\right)$ versus $\Re \check{z}\left(  t\right)$  for irrational values of $q$ and $0\leq t\leq 8 \pi$.}
\label{time_evolution-q1}
\end{figure}
\begin{figure}[t]
\centering
 \subfigure [
$q=\pi$.  \hskip5cm (b)  ]{
\includegraphics[width=0.45\textwidth]{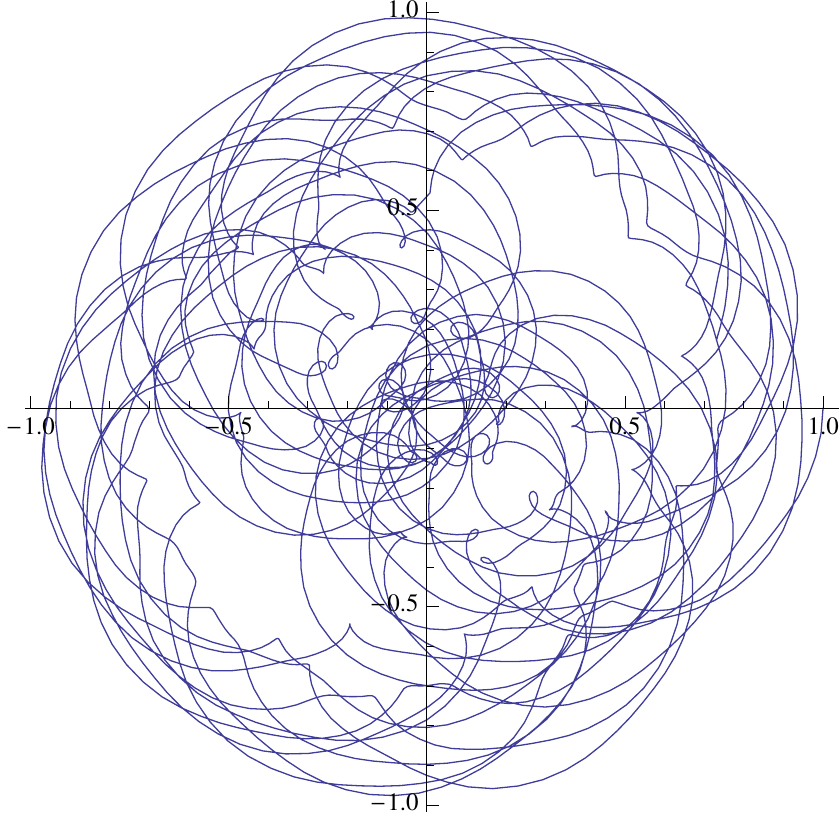}\qquad  
\includegraphics[width=0.45\textwidth]{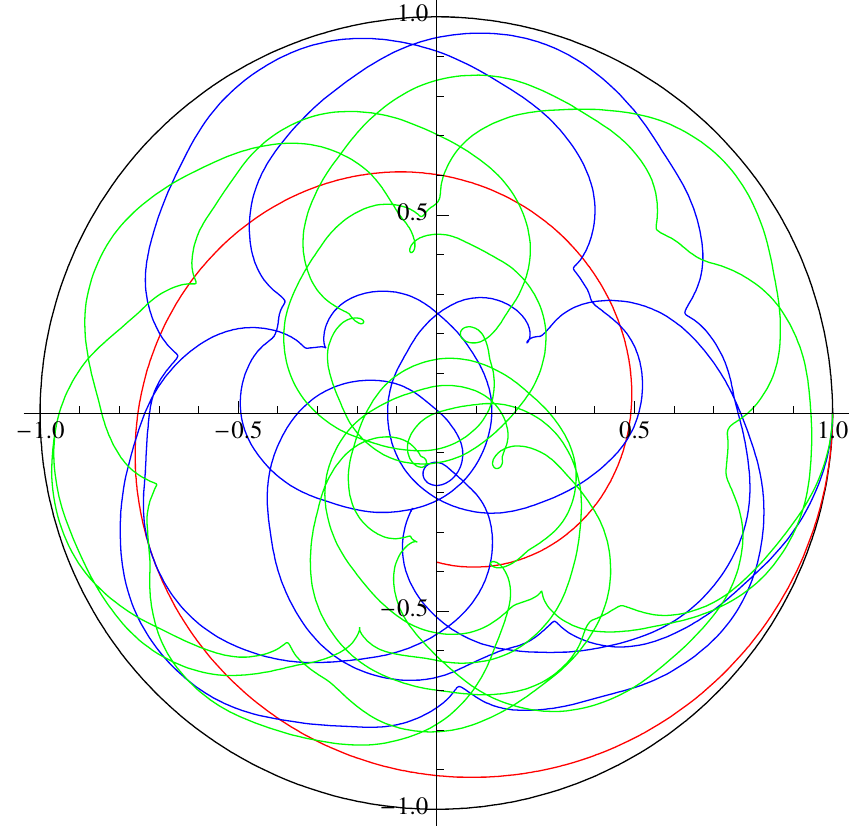}
}
\caption{\footnotesize Plots of ${\Im} \check{z}\left(  t\right)$ versus $\Re \check{z}\left(  t\right)$  for irrational values of $q$ and $0\leq t\leq 8 \pi$. In Fig.~\ref{time_evolutions3}(b) we plot  the previous three figures together with the corresponding to $q=1$ for   
$0\leq t\leq 2 \pi$.}
\label{time_evolutions3}
\end{figure}

In this paper we have introduced a $q$-dependent family of coherent states and we have explored some properties of  the quantum harmonic oscillator obtained through the corresponding coherent state quantization. We have restricted our study to the case in which $q^{-1}$ is a quadratic unit Pisot number, since then the  spectrum of the quantum Hamiltonian is made of the $q$-deformed integers ${}^{[s]}[n]_{q} = \frac{q^n - q^{-n}}{q-q^{-1}}$  which are still integers and form  sequences of  Fibonacci type.   We have  put into evidence  interesting quantum features  issued from these particular algebraic cases, concerning particularly  the  localization in the configuration space and in the phase space,  probability distributions and related statistical features,  time evolution  and  semi-classical phase space trajectories.  The periodicity of the latter nicely reflects the algebraic Pisot nature of the deformation parameter $q$. By contrast  we present the semi-classical phase space trajectories for irrational values of $q$ (Figs.~\ref{time_evolution-q1} and Fig.~\ref{time_evolutions3}). Obviously, the trajectories are not periodic.

Let us place the present study into a more general perspective. ``Coherent state(s) quantization'' \cite{gazeau09} is a generic phrase for naming a certain point of view in analyzing a set $X$ (here the complex plane) of parameters,  equipped with a measure $\mu$ (here $w_q(\vert z\vert^2)\,d^2z/\pi$). The approach matches  what physicists  understand by quantization when the ``observed'' measure space $X$ has a phase space or symplectic structure (it is not exactly the case here since the measure is not the usual symplectic one).  It matches also well established  approaches by signal analysts, like wavelet analysis \cite{L1}.   The set $X$ can be finite, countably infinite or uncountably infinite. The approach is generically simple, of Hilbertian essence,  and always the same: one starts from the Hilbert space $L^2(X,\mu)$ of complex square integrable functions on $X$ with respect to the measure $\mu$. One chooses an orthonormal set $\mathcal{O}$ of vectors $\phi_n$ in it (here $\phi_n(x) = \bar x^n/x_n!$ satisfying the finiteness condition  $\mathcal{N}(x)= \sum_n \vert\phi_n(x)\vert^2 < \infty$, and a ``companion" Hilbert space $\mathcal{H}$ (the space of ``quantum states'')  with orthonormal basis (here the $|e_n\rg$'s) in one-to-one correspondence with the elements of  $\mathcal{O}$. There results a family $\mathcal{C}$ of states 
\begin{equation*}
\label{gencs}
|x\rangle = \frac{1}{\sqrt{\mathcal{N}(x)}}\sum_n \overline{\phi_n(x)}\,|e_n\rg
\end{equation*}
 (the ``coherent states'') in $\mathcal{H}$, which are labelled by elements of $X$ and which resolve the unity operator in $\mathcal{H}$. This is the departure point for analysing the original set and functions living on it from the point of view of the frame (in its true sense) $\mathcal{C}$. We end in general with a non-commutative algebra of operators in $\mathcal{H}$, set aside the usual questions of domains in the infinite dimensional case.  

There is a kind of manifest universality in this approach. The change of the frame family $\mathcal{C}$ produces another quantization, another point of view, possibly equivalent to the previous one, possibly not. The present study  lies in the continuity of a series of such explorations, which were already present in the first works by Klauder at the beginning of the sixties  of the past century (see for instance \cite{klauder95,main:klauderpath} and references therein), pursued by Berezin \cite{berezin75} in his famous paper of 1975, and more recently extended to various measure sets (see for instance \cite{CS44,CS52,CS47,CS57,cotgavour11,CS61}).

That this correspondence fits the CS quantization is once more the mark of the universality and the easy implementation of this type of analysis, in comparison with usual quantization methods  (see the review \cite{aliengl05}).



\appendix

\section*{Appendix}

\subsection*{A\;\; Inequalities}
\label{ineq}
Let us consider the Hilbert space $L^2([0, R), w(t) dt)$ of square integrable functions on the interval $[0,R)\subset\R$ with respect the measure $w(t)\; dt$. Scalar product and norms are  respectively defined by
\begin{equation*}
\lg f_1 | f_2\rg = \int_0^R \overline{f_1}(t)\, f_2(t)\,w(t)\, dt\, , \qquad \Vert f\Vert = \sqrt{\int_0^R \vert f(t)\vert^2\,w(t)\, dt}\, .
\end{equation*}
In particular for the monomial functions $m_{\alpha}(t) := t^{\alpha}$, 
\begin{equation*}
\left\Vert m_{\frac{n}{2}}\right\Vert^2  =  \int_0^R t^n \,w(t)\, dt = x_n!\, . 
\end{equation*}
From the Cauchy-Schwarz inequality, $\vert \lg f_1 | f_2\rg\vert \leq \Vert f_1\Vert\, \Vert f_2\Vert$   valid for any pair $f_1,f_2,$  in $L^2([0, R), w(t) dt)$
we infer the inequality:
\begin{equation}
\label{factineq}
x_{\frac{n_1+n_2}{2}} ! = \int_0^R t^{\frac{n_1+n_2}{2}} \,w(t)\, dt =  \left\lg m_{\frac{n_1}{2}} 
| m_{\frac{n_2}{2}} \right\rg \leq \left\Vert m_{\frac{n_1}{2}} \right\Vert\,\cdot 
\left\Vert m_{\frac{n_2}{2}}\right\Vert
= \sqrt{x_{n_1}!\,x_{n_2}!}\, ,
\end{equation}
for any $n_1,n_2\in \N$, actually for any $n_1,n_2 \in \R^+$. 

Now, consider the series defined for $k\in \N$ and for $t\geq 0$ by:
\begin{equation*}
\label{dmrA}
\mathcal{S}_k(t)= \frac{1}{\mathcal{N}(t)}\sum_{n=0}^{\infty} \frac{x_{\frac{k}{2}+n}!}{x_n! x_{n+k}!}\,t^{n + k/2}\,.
\end{equation*}
Due to (\ref{factineq}) we have a first upper bound:
\begin{equation*}
\label{dmrAineq1}
\mathcal{S}_k(t)= \frac{1}{\mathcal{N}(t)}\sum_{n=0}^{\infty} \frac{x_{\frac{k}{2}+n}!}{\sqrt{x_n! x_{n+k}!}}\, \frac{1}{\sqrt{x_n! x_{n+k}!}}\,t^{n+k/2}\leq  \frac{1}{\mathcal{N}(t)}\sum_{n=0}^{\infty} \frac{1}{\sqrt{x_n! x_{n+k}!}}\,t^{n+k/2}\, .
\end{equation*}
Let us apply again the Cauchy-Schwarz inequality:
\begin{equation*}
\label{dmrAineq2}
\sum_{n=0}^{\infty} \frac{1}{\sqrt{x_n! x_{n+k}!}}\,t^{n+k/2} \leq \sqrt{\sum_{n=0}^{\infty} \frac{1}{x_n! }\,t^{n}}\, \sqrt{\sum_{n=0}^{\infty} \frac{1}{x_{n+k}!}\,t^{n+k}} \leq \mathcal{N}(t)\, .
\end{equation*}
In consequence, we can assert that
\begin{equation*}
\label{dmrAineq3}
\mathcal{S}_k(t) \leq 1\qquad 
\forall\ k\geq 0\, , \, t\geq 0 \,.
\end{equation*}

\subsection*{B\;\; Standard $q$-calculus}
\label{qcal}

There exists a large literature  \cite{thomae,jackson_1904,koornwinder_92,exton,gasper_rahman,aar}, devoted to $q$-calculus in the standard sense (see specially  \cite{solkac} for updated results), which means that the $q$-deformation of a number $x$ is given by the asymmetric expression
\begin{equation*}
\label{asqdef}
[x]_q := \frac{1-q^x}{1-q}\, . 
\end{equation*}
Suppose here and further that $0<q<1$ and let us adopt the following notations for the related infinite products
\begin{eqnarray}
 \label{intr2}
(a+b)_q^n &=& \prod_{j=0}^{n-1}(a+q^jb)\ ,\quad \text{ if } n\in\Z^+\ ,
\\  \label{intr3}
(1+a)_q^\infty &=& \prod_{j=0}^{\infty}(1+q^ja)\ ,
\\
\label{intr4}
(1+a)_q^t &=& \frac{(1+a)_q^\infty}{(1+q^ta)_q^\infty}\ ,
 \quad \text{ if } t\in\C\ . 
\end{eqnarray}
Under the assumptions on $q$, the infinite product (\ref{intr3})
is convergent, and the definitions (\ref{intr2}) and (\ref{intr4}) 
are consistent.

The $q$--gamma function $\G_q(t)$, a $q$--analogue of Euler's gamma function,
was introduced by Thomae \cite{thomae} and later by Jackson 
\cite{jackson_1904} as the infinite product
\begin{equation*}\label{int1}
\G_q(t)=\frac{(1-q)_q^{t-1}}{(1-q)^{t-1}}\ ,\quad t>0\ .
\end{equation*}
A first   integral representation of $\G_q(t)$ reads as
\begin{equation}\label{intr5}
\G_q(t)=\int_0^{\frac{1}{1-q}}x^{t-1}E_q^{-qx}d_qx\ .
\end{equation}
Here in \eqref{intr5} $ E_q^x$ is one of the two $q$--analogues of the exponential function,
\begin{eqnarray}
E_q^x &=& \sum_{n=0}^\infty q^{n(n-1)/2}\frac{x^n}{[n]!}\ 
=\ (1+(1-q)x)_q^\infty\ , \label{intr6} \\
e_q^x &=& \sum_{n=0}^\infty \frac{x^n}{[n]!}\ 
=\ \frac{1}{(1-(1-q)x)_q^\infty}\ , \label{intr7}
\end{eqnarray}
and the $q$--integral (introduced by Thomae \cite{thomae} 
and Jackson \cite{jackson_1910}) is defined by
\begin{equation}\label{intr8}
\int_0^{a}\; f(x)\; d_qx
=(1-q)\sum_{j=0}^\infty a\; q^jf(a\; q^j)\ .
\end{equation}
Notice that the series on the right--hand side of  \eqref{intr8} is guaranteed to be convergent as soon as the function $f$ is such that, for some $C>0,\ \alpha>-1$,
$|f(x)|<Cx^{\alpha}$ in a right neighborhood of $x=0$.
Also note that the $q$--exponential functions \eqref{intr6}-\eqref{intr7} are related by
\begin{equation*}
\label{eErel}
e_q^x\;E_q^{-x}=E_q^x\;e_q^{-x}=1\, ,
\end{equation*}
and  that for $q\in(0,1)$ the series expansion of $e_q^x$ has 
radius of convergence $1/(1-q)$, whereas the series expansion of $E_q^x$ converges for every $x$.

The $q$--derivative $D_q$ of a function is defined by 
$$
(D_q f)(x)=\frac{f(qx)-f(x)}{(q-1)x}\ .
$$
Jackson integral and $q$--derivative are related by the ``fundamental
theorem of quantum calculus''  \cite[p.~73]{kac_qc}, i.e. 
\begin{theorem}
\begin{itemize}
\item[a)] If $F$ is any anti $q$--derivative of the function $f$, 
namely $D_qF=f$, continuous at $x=0$, then
$$
\int_0^af(x)d_qx=F(a)-F(0)\ .
$$
\item[b)] For any function $f$ one has
$$
D_q\int_0^xf(t)d_qt=f(x)\ .
$$
\end{itemize}
\end{theorem}
The  $q$--analogue of the Leibniz rule is, as it is easy to check,
\begin{equation}\label{qleibniz}
D_q(f(x)g(x))=g(x)D_qf(x)+f(qx)D_qg(x)\ .
\end{equation}
An immediate consequence is the $q$--analogue 
of the rule of integration by parts:
$$
\int_0^ag(x)D_qf(x)d_qx
=f(x)g(x)\Big|_0^a-\int_0^af(qx)D_qg(x)d_qx\ .
$$
The Jackson integral in a generic interval $[a,b]$ is defined by 
\cite{jackson_1910}:
$$
\int_a^bf(x)d_qx=\int_0^bf(x)d_qx-\int_0^af(x)d_qx\ .
$$
Improper integrals are also defined in the following way
\cite{jackson_1910}, \cite{koornwinder_99}:
\begin{equation}\label{improp}
\int_0^{\infty/A}f(x)d_qx
=(1-q)\sum_{n\in\Z}\frac{q^n}{A}f\Big(\frac{q^n}{A}\Big)\ .
\end{equation}
Notice that in order the series on the right--hand side  of \eqref{improp} to be convergent,
it suffices that the function $f$ satisfies the conditions:
$|f(x)|<Cx^\alpha, \quad \forall x\in [0,\epsilon)$, 
for some $C>0,\ \alpha>-1,\ \epsilon>0$;
and $|f(x)|<Dx^\beta,\quad\forall x\in [N,\infty)$, 
for some $D>0,\ \beta<-1,\ N>0$.
In general though, even when these conditions are satisfied,
the value of the sum will be dependent on the constant $A$.
In order the integral to be independent of $A$, the anti $q$--derivative
of $f$ needs to have limits for $x\rightarrow 0$ and 
$x\rightarrow+\infty$.

The $q$--exponential functions \eqref{intr6}-\eqref{intr7} satisfy the following properties:
\begin{enumerate}
\item $D_qe_q^x=e_q^x\ ,\qquad D_qE_q^x=E_q^{qx}$.
\item $e_q^x\;E_q^{-x}=E_q^x\;e_q^{-x}=1$.
\item $e_q^x=E_{1/q}^x$.
\end{enumerate}

The function $\G_q(t)$ is the ``correct'' $q$--analogue of the
$\Gamma$--function, since it reduces to $\G(t)$ in the limit $q\rightarrow1$, and it satisfies a property analogue to $\G(t+1) = t\, \G(t)$. Indeed, $\G_q(t)$ is equivalently expressed as 
\begin{equation*}\label{a}
\G_q(t)=\frac{(1-q)_q^{t-1}}{(1-q)^{t-1}}\ ,
\end{equation*}
and, in particular, it verifies
\begin{equation*}
\label{fcteqGq}
\G_q(t+1)=[t]\;\G_q(t)\ ,\quad \forall t>0\ ,\qquad \G_q(1)=1\ .
\end{equation*}

\subsection*{C\;\; The essential of symmetric $q$-calculus}

We now recall the essential of the $q$-calculus in the symmetric case, i.e. when the $q$-deformation of a number, say $t$, is given by 
\begin{equation*}
\label{qsdeft}
{}^{[s]}[t]_{q} = \frac{q^t - q^{-t}}{q-q^{-1}}\, .
\end{equation*}
To simplify, we will drop the subscript $q$ in all symbols when it is not really needed for the understanding.

\paragraph{$q$-derivative :}
the symmetric $q$-derivative is defined as
\begin{equation*}
\label{qsderiv}
{}^{[s]}D_q f(x)= \frac{f(qx) - f(q^{-1}x)}{(q-q^{-1})x}\, .
\end{equation*}
 Applied to a power of the variable $x$, it gives:
\begin{equation*}
\label{derx}
\sd x^n = \qn x^{n-1}\, .
\end{equation*}

The   Leibnitz formula \eqref{qleibniz} reads in the present context:
\begin{equation*}
\label{qleibnitz}
\sd (f(x)g(x)) = f(q^{-1}x) \sd g(x) + g(qx) \sd f(x)\, .
\end{equation*}

\paragraph{``Proper'' definite $q$-integral :}
\begin{equation}
\label{qsint}
\int_0^a f(x) \, \sm x = (1-q^2) a \sum_{n=0}^{\infty} q^{2n}f(q^{2n+1}a)\, .
\end{equation}
Applied to a power of the variable $x$, it gives:
 \begin{equation*}
\label{intx}
\int_0^a x^n \, \sm x = \frac{1}{\qnp} a^{n+1}\, .
\end{equation*}

If a $q$-primitive $F(x)$ of $f(x)$ is known (i.e. $f(x) = \sd F(x)$), then the integral (\ref{qsint}) is, as expected,  equal to
\begin{equation*}
\label{prqsint}
\int_0^a f(x) \, \sm x = F(a) - F(0)\, .
\end{equation*}

\paragraph{``Improper'' definite $q$-integral :}
\begin{equation*}
\label{iqsint}
\int_0^{\infty/A} f(x) \, \sm x = (1-q^2)  \sum_{n\in \Z} \frac{q^{2n}}{A}f\left(\frac{q^{2n+1}}{A}\right)\, .
\end{equation*}
If $f(x) = \sd F(x)$, then this integral is equal to 
\begin{equation*}
\label{priqsint}
\int_0^{\infty/A} f(x) \, \sm x = \underset{N \to \infty}{\lim}
\left( F\left(\frac{q^{-2N}}{A}\right) - F\left(\frac{q^{2N}}{A}\right)\right)\, .
\end{equation*}

\paragraph{Factorials and  
$q$-exponentials :}
the (symmetric) $q$-factorial is defined by
\begin{equation*}
\label{qfact}
\qn! = \qn \times   \qnm \times \cdots \times {}^{[s]}[2]\times 1\, .
\end{equation*}

In the context of $q$-deformations, there exist two types of $q$-exponentials:
\begin{eqnarray*}
\label{qexp}
 \mathfrak{e}_q(x)   &= \sum_{n=0}^{\infty} \frac{x^n}{\qn!}\, ,   \\[0.3cm]
\label{qExp} \mathfrak{E}_q (x)   &=  \sum_{n=0}^{\infty} 
q^{\frac{n(n+1)}{2}}\frac{x^n}{\qn!} \, .
\end{eqnarray*}
First we note that the  symmetries 
\begin{equation*}
\label{symexp}
 \mathfrak{e}_{1/q}(x) =  \mathfrak{e}_q(x)  \, , \qquad 
 \mathfrak{E}_{1/q} (x)=
 1/ \mathfrak{E}_q(-x) \, .
 \end{equation*}
These two series obey the $q$-differential equations:
\begin{equation*}
\label{difexp}
\sd\, \mfe_q(x) = \mfe_q(x)\, , \qquad \sd\, \mfE_q(x) = \mfE_q(qx)\, .
\end{equation*}

\paragraph{$q$--Gamma function :}

we define a  gamma-type function through the proper definite integral:
\begin{equation*}
\label{firstgam}
{}^{[s]}\widetilde{\gamma}_q(t) = \int_0^{\frac{1}{1-q^2}} x^{t-1}\, \mfE_q(-qx)\, \sm x\, .
\end{equation*}
 By performing an integration by parts, we can prove the following recurrence relation:
 \begin{equation*}
\label{recgam}
{}^{[s]}\widetilde{\gamma}_q(t + 1) = q^t\, {}^{[s]}[t] \, {}^{[s]}\widetilde{\gamma}_q(t)\, .
\end{equation*}
In particular, for $t = n \in  \N$, we get:
 \begin{equation*}
\label{recgam1}
{}^{[s]}\widetilde{\gamma}_q(n+ 1) = q^{\frac{n(n+1)}{2}}\, {}^{[s]}[n]! \, .
\end{equation*}


\subsection*{D\;\; Equation and solutions for general quadratic Pisot numbers}

We consider two cases:

\subsubsection*{D.1\;\; Positive conjugate case}
Equation:
\begin{equation*}
\label{algPis+}
X^2 - (a+1)X + (a-b) = 0; \qquad a, b \in \mathbb{Z}, \;\; 
a \geq b+1, \;\; b > 0.
\end{equation*}
Solutions:
\begin{equation*}\begin{array}{l}
\label{sol+}
1< a <\beta = \frac{1}{2}\left(a+1 + \sqrt{(a-1)^2 + 4b}\right), \\[0.3cm]  0< \beta' = \frac{1}{2}\left(a+1 - \sqrt{(a-1)^2 + 4b}\right) <1,
\end{array}\end{equation*}
with
\begin{equation*}
\label{propsol+}
\lfloor \beta \rfloor = a, \qquad
 \beta \beta' = a-b \geq 1, \qquad
  \beta + \beta' = a +1.
\end{equation*}

\subsubsection*{D.2\;\; Negative conjugate case}
Equation:
\begin{equation*}
\label{algPis-}
X^2 - cX - d = 0; \qquad  c, d \in \mathbb{Z}, \;\; c \geq d \geq 1.
\end{equation*}
Solutions:
\begin{equation*}
\label{sol-}
1 \leq c < \beta = \frac{1}{2}\left(c + \sqrt{c^2 + 4d}\right), \qquad 
 -1 < \beta' = \frac{1}{2}\left(c - \sqrt{c^2 + 4d}\right) < 0,
\end{equation*}
with
\begin{equation*}
\label{propsol-}
\lfloor \beta \rfloor = c, \qquad
  \beta \beta' = -d \geq 1, \qquad
   \beta + \beta' = c.
\end{equation*}


\subsection*{E\;\; Powers and recurrence for general quadratic Pisot numbers}

\subsubsection*{E.1\;\; Positive conjugate case}
\begin{equation*}
\label{pow+}
\beta^n = v_{n} \beta  + w_{n}, \quad   v_0 = 0, \;\; w_0= 1,\quad 
v_1 = 1, \;\;  w_1 = 0.
\end{equation*}
\begin{equation*}
\label{rec+}
w_{n+1} = (b-a) v_{n} ,  \ v_{n+1}= (a + 1)v_{n}  + (b-a)v_{n-1}.
\end{equation*}
\begin{equation*}
\label{difpow+}
v_{n} = \frac{\beta^n - \beta'^n}{\beta - \beta'}
\end{equation*}
\begin{equation*}
\label{sumpow+}
\beta^n + \beta'^n = (a+1)v_{n} + 2 (b-a) v_{n-1}.
\end{equation*}

\subsubsection*{E.1\;\; Negative conjugate case}
\begin{equation*}
\label{pow-}
\beta^n = v_{n} \beta  + w_{n}, \   v_0 = 0, \ w_0= 1, v_1 = 1, 
\  w_1 = 0.
\end{equation*}
\begin{equation*}
\label{rec-}
w_{n+1} = dv_{n},  \ v_{n+1}= cv_{n}  + d v_{n-1}.
\end{equation*}
\begin{equation*}
\label{difpow-}
v_{n} = \frac{\beta^n - \beta'^n}{\beta - \beta'}
\end{equation*}
\begin{equation*}
\label{sumpow-}
\beta^n + \beta'^n = c v_{n} + 2d v_{-1}.
\end{equation*}


\section*{Acknowledgments}

This work was partially supported  by the Ministerio de Educaci\'on
y Ciencia  of Spain (Project FIS2009-09002). J-P. G. acknowledges
the University of Valladolid  for its hospitality and financial support.



\end{document}